\documentclass[prd, reprint, lengthcheck, floatfix, superscriptaddress]{revtex4-1}

\usepackage{amsmath}
\usepackage{amssymb}
\usepackage{mathtools}
\usepackage{bm}
\usepackage{graphicx,color}
\usepackage{hyperref}
\usepackage{braket}
\usepackage{textcomp}
\usepackage{tikz}
\usepackage{ragged2e}
\usepackage[utf8]{inputenc}
\usepackage{wrapfig}
\usepackage{booktabs}
\usepackage[english]{babel}
\usepackage{epstopdf}

\mathtoolsset{showonlyrefs}

\hypersetup{
  colorlinks   = true, 
  urlcolor     = blue, 
  linkcolor    = blue, 
  citecolor   = red, 
}

\newcommand{\tr}{\operatorname{tr}}

\newcommand{\ic}{\ensuremath{\mathrm{i}}}
\newcommand{\dee}{\ensuremath{\mathrm{d}}}
\renewcommand{\vec}[1]{\ensuremath{\bm{#1}}}

\begin{document}

\title{Matrix product states and variational methods applied to critical quantum field theory}


\author{Ashley Milsted}
\email[]{ashley.milsted@itp.uni-hannover.de}
\affiliation{Leibniz Universit\"at Hannover, Institute of Theoretical Physics, Appelstrasse 2, D-30167 Hannover, Germany}
\author{Jutho Haegeman}
\affiliation{Vienna Center for Quantum Science and Technology, Faculty of Physics, University of Vienna, Boltzmanngasse 5, A-1090 Wien, Austria}
\affiliation{Faculty of Physics and Astronomy, University of Ghent, Krijgslaan 281 S9, 9000 Gent, Belgium}
\author{Tobias J.\ Osborne }
\affiliation{Leibniz Universit\"at Hannover, Institute of Theoretical Physics, Appelstrasse 2, D-30167 Hannover, Germany}

\date{\today}

\begin{abstract}
  We study the second-order quantum phase-transition of massive real scalar field theory 
  with a quartic interaction in (1+1) dimensions on an infinite spatial lattice using 
  matrix product states (MPS).
  We introduce and apply a naive variational conjugate gradient method, based on the
  time-dependent variational principle (TDVP) for imaginary time, to obtain
  approximate ground states, using a related ansatz for excitations to 
  calculate the particle and soliton masses and to obtain the spectral 
  density. We also estimate the central charge using finite-entanglement scaling.
  Our value for the critical parameter agrees well with recent Monte Carlo
  results, improving on an earlier study which used the related DMRG method,
  verifying that these techniques are well-suited to studying critical field systems.
  We also obtain critical exponents that agree, as expected, 
  with those of the transverse Ising model.
  Additionally, we treat the special case of uniform
  product states (mean field theory) separately, showing that they
  may be used to investigate non-critical quantum field theories under certain
  conditions.
\end{abstract}

\maketitle

\section*{Introduction}
\label{chap:intro}

Quantum field theories (QFT) \cite{peskin_introduction_1995} are extremely
good at describing and predicting the behavior of fundamental particles, 
as demonstrated by the prediction of the Higgs boson over forty years ago and 
its recent apparent discovery \cite{atlas_observation_2012}. 
Very often, however, obtaining predictions from QFT is difficult
due, in no small part, to the huge Hilbert spaces they are set in.
In many cases, such as quantum electrodynamics (QED), perturbation theory has been 
used very successfully, yet some important phenomena are not accessible to 
these methods, notably confinement in quantum chromodynamics (QCD) \cite{creutz_quarks_1985}. 
Lattice regularizations of QFT's have been 
very useful in such cases, often in combination with 
Monte Carlo numerical techniques. Here, however,
the sign problem \cite{von_der_linden_quantum_1992} 
presents a challenge - not to mention that such simulations often require very
large computational resources to produce useful results (see \cite{durr_ab_2008},
where hadron masses are determined with the help of clusters and a supercomputer). 

Meanwhile, the numerical study of lattice 
systems in one spatial dimension has benefited greatly from the density matrix 
normalization group (DMRG) method,
which has been recognized as a variational method producing approximate
ground states that are matrix product states (MPS)
\cite{white_density_1992, schollwock_density-matrix_2005}. 
There is a direct link between the dimension of the
MPS parameter-space and the amount of entanglement that a state can contain,
allowing efficient representation of a great many relevant states 
\cite{eisert_colloquium:_2010, hastings_area_2007, osborne_efficient_2006}, 
in particular ground states and low-lying excited states of gapped systems
\cite{de_beaudrap_solving_2010, hastings_solving_2006, masanes_area_2009},
which all lie in the low-entanglement ``corner'' of Hilbert space. 
Recently, other variational techniques have been applied to MPS such as 
the time-dependent variational principle (TDVP) \cite{haegeman_time-dependent_2011},
which permits efficient simulation of dynamics, and a related excitation ansatz 
for the determination of dispersion relations \cite{haegeman_variational_2012} 
for translation-invariant systems in the thermodynamic limit.

Given the great successes of MPS and variational methods in studying lattice 
systems, it is natural to ask whether they can be usefully applied to lattice quantum 
fields in $(1+1)$ dimensions, and whether continuum results can be extracted 
efficiently.
A useful test case is $\phi^4$-theory which, despite its simplicity, exhibits
interesting behavior such as spontaneous symmetry-breaking. 
It contains a second-order quantum phase-transition
in $(1+1)$ dimensions \cite{chang_existence_1976} and is expected to belong to 
the same universality class as the transverse Ising model \cite{simon_phi4_2_1973} 
so that critical exponents should be the same for both.
In fact, DMRG has already shown promise when applied to
$\phi^4$-theory \cite{sugihara_density_2004, weir_studying_2010}, reproducing the expected 
critical behavior and obtaining values of the critical parameter close to
those of Monte Carlo studies \cite{loinaz_monte_1998, schaich_improved_2009, de_investigations_2005}.

We use variational methods with MPS to obtain the ground-state field expectation 
value and low-lying excitation energies of $\phi^4$-theory.
The scaling of these quantities in parameter-space allows us to locate the 
critical point and to determine the critical exponents. 
We begin by introducing QFT and real scalar field
theory in section \ref{chap:QFT}, showing how it can be put on a spatial lattice
and discussing its critical behavior. In section \ref{chap:MPS}, we define
the uMPS variational class and the corresponding TDVP algorithm and excitation ansatz
before detailing our variational conjugate-gradient method for finding ground states.
Section \ref{chap:phi4} is the main part of this work, in which we apply these
techniques to $\phi^4$-theory and obtain our estimate for the continuum 
critical parameter, which we compare with previous
results from the literature, as well as values for critical exponents and the
central charge. We also separately assess the usefulness of 
mean-field theory (a special case of MPS) for studying QFT, which is 
an attractive tool because of the low computational complexity required to 
estimate physical quantities.

We have kept the software developed for this work intentionally general such that
it may be of use to others. It is available under a permissive open-source 
license \cite{milsted_evomps_????}.

\section{Quantum field theory}
\label{chap:QFT}
We introduce the basic principles of quantum field theory using the
Hamiltonian formulation with real scalar fields as an
example, defining interacting $\phi^4$-theory both in the continuum 
and on a spatial lattice. We then discuss its spontaneous 
symmetry-breaking, which corresponds to a second-order quantum
phase-transition in $(1+1)$ dimensions. The beginning of this section is based
partly on lectures on quantum field theory given by Marco Zagermann at Leibniz 
Universität Hannover in 2010/11 and also on \cite{peskin_introduction_1995}.

\subsection{Real scalar field without interactions}

Quantum fields in Minkowski space-time are quantum systems set in an uncountably 
infinite-dimensional Hilbert space which can usually be divided naturally into 
subsystems corresponding to points in momentum space. They can often be constructed from 
a corresponding classical field defined by a Lorentz-invariant action. 
The classical field is then quantized in such a way as to produce a consistent 
Hilbert space and Hamiltonian where Lorentz-invariance and causality are maintained.

As an example, take a classical real scalar field $\phi(x) \in \mathbb{R}$ with action
\begin{align}
  S = \int \dee x \underbrace{\frac{1}{2}(\partial_\nu \phi \partial^\nu \phi 
        - \mu_0^2 \phi^2)}_{\mathcal{L}},
 \label{eq:phi_free_action}
\end{align}
where $\nu = 0 \dots d - 1$, $\partial_\nu = \partial / \partial x^\nu$
and scalar products are defined via the Minkowski metric with 
the ``mostly-minus'' signature $(1, -1 \dots -1)$. We use $\vec{x}$ to denote
the spatial part of a Minkowski vector $x$. The integrand $\mathcal{L}[\phi(x), \partial \phi(x)]$
is called the Lagrangian density.
Stability with respect to a small variation $\delta\phi$ requires
that the Euler-Lagrange equations
\begin{align*}
  \frac{\partial \mathcal{L}}{\partial \phi} 
  - \partial_\nu \left( \frac{\partial \mathcal{L}}{\partial (\partial_\nu \phi)} \right)
  = 0
\end{align*}
are satisfied, leading in this case to the equation of motion
\begin{align*}
  (\Box + \mu_0^2) \phi(x) = 0,
\end{align*}
where $\Box = \partial_\nu \partial^\nu$, which is the Klein-Gordon equation.
The lack of non-linear field-terms in the equations of motion makes this a free 
(non-interacting) field.
Performing a Fourier transform
$\phi(t, \vec{x}) = \int \frac{\dee \vec{p}}{(2\pi)^{d-1}} e^{\ic \vec{p}.\vec{x}} \phi(t, \vec{p})$,
we can rewrite the equations of motion as
\begin{align*}
  (\partial_t^2 + \vec{p}^2 + \mu_0^2) \phi(t, \vec{p}) = 0,
\end{align*}
which has the form of a simple harmonic oscillator with angular
frequency $\omega(\vec{p}) = E(\vec{p}) = \sqrt{\vec{p}^2 + \mu_0^2}$. We can thus think
of the classical Klein-Gordon field as a set of independent harmonic oscillators, 
one for each point in momentum space. 

To quantize this free scalar field theory, we start from the classical Hamiltonian.
The canonical conjugate momentum $\pi(x)$ corresponding to the coordinate 
$\phi(x)$ is
\begin{align*}
  \pi(x) = \frac{\partial \mathcal{L}}{\partial(\partial_0 \phi(x))} = \partial_0 \phi(x) = \dot{\phi}(x)
\end{align*}
and, performing a Legendre transformation, the Hamiltonian density is
\begin{align*}
  \mathcal{H} = \pi \dot{\phi} - \mathcal{L} = 
              \frac{1}{2}(\pi^2 + (\nabla \phi)^2 + \mu_0^2 \phi^2).
\end{align*}
The coordinate $\phi(x)$ and the momentum $\pi(x)$ obey the Poisson-bracket
relationship
\begin{align*}
  \left\{ \phi(t, \vec{x}), \pi(t, \vec{y}) \right\} = \delta (\vec{x} - \vec{y}),
\end{align*}
where $\delta(\vec{x})$ is the $(d-1)$-dimensional Dirac delta distribution. 
The ingredients required for a canonical quantization of the classical theory
are now ready. To proceed, we replace the classical phase-space coordinates
in the above relations
with operators (one for each space-time coordinate) obeying the commutation relation
\begin{align*}
  \left[ \phi(t, \vec{x}), \pi(t, \vec{y}) \right] = \ic \delta (\vec{x} - \vec{y}).
\end{align*}
The field operator $\phi = \phi^\dagger$ is Hermitian because the classical field was 
real-valued. We now have an operator-valued field where 
the operators $\phi(x)$ must obey the Klein-Gordon equation.
The harmonic oscillator picture of the classical field suggests attempting to 
write solutions in terms of quantum harmonic oscillators. 
Making the same move to momentum space as before, we can 
write a general solution as a superposition of plane waves using Fock-space 
creation and annihilation operators $a_{\vec{p}}^\dagger$ and  $a_{\vec{p}}$
\begin{align*}
  \phi(x) = \int \frac{\dee \vec{p}}{(2\pi)^{d-1}} \frac{1}{\sqrt{2 E(\vec{p})}}
            \left( a_{\vec{p}} e^{-\ic p\cdot x} + a^\dagger_{\vec{p}} e^{\ic p\cdot x} \right)|_{p^0 = E(\vec{p})},
\end{align*}
where $[a_{\vec{p}}, a^\dagger_{\vec{q}}] = (2\pi)^{d-1} \delta (\vec{p} - \vec{q})$
and $p^0 = E(\vec{p}) = \sqrt{\vec{p}^2 + \mu_0^2}$. The Hilbert space contains
the vacuum $a_{\vec{p}}\ket{0} = 0 \quad \forall \vec{p}$ and countably infinite
excited states $(a_{\vec{p}}^\dagger)^n \ket{0}$ for each momentum 
$\vec{p} \in \mathbb{R}^{d-1}$.
Using $\pi(x) = \dot{\phi}(x)$, we can write the Hamiltonian as
\begin{align*}
  H = \int \frac{\dee \vec{p}}{(2\pi)^{d-1}} E(\vec{p})
      \left( a_{\vec{p}}^\dagger a_{\vec{p}} + \frac{1}{2} [a_{\vec{p}}, a_{\vec{p}}^\dagger] \right),
\end{align*}
where the second term does not annihilate the vacuum, leading to an infinite
vacuum energy contribution. This is perhaps not too surprising:
We are summing up an infinite number of ground state energies,
one for each Fourier mode, each of which is the energy
contained within an infinite volume of space.
Since it is energy differences that 
are observable, and not absolute energies, this infinite contribution 
should not cause any problems. 
A general eigenstate 
$a_{\vec{p}}^\dagger a_{\vec{q}}^\dagger \dots \ket{0}$ has energy 
(ignoring the infinite vacuum contribution) $E(\vec{p}) + E(\vec{q}) + \dots$
and is also a momentum eigenstate with momentum $\vec{p} + \vec{q} + \dots$, 
where the momentum operator can be obtained via the classical theory as the 
conserved quantity associated with spatial translations (using Noether's 
theorem). The field operator $\phi(t, \vec{x})$ (in the Heisenberg picture) acts on the 
vacuum to create a superposition of momentum eigenstates resulting in a particle 
localized at the space-time coordinate $x$.

A quantity that turns out to be very useful is the two-point correlation
function $\braket{0|\phi(x)\phi(y)|0}$, which can be interpreted as the 
probability of a particle created at point $x$ propagating to point $y$ (or
vice-versa, depending on the time-coordinates). 
For this reason it is also called the propagator. It has the form
\begin{align*}
  D(x - y) &= \braket{0|\phi(x)\phi(y)|0}  \\
    &= \int \frac{\dee \vec{p}}{(2\pi)^{d-1}} \frac{1}{2E(\vec{p})} e^{-\ic p\cdot(x-y)} |_{p^0 = E(\vec{p})}.
\end{align*}
A related quantity is the Feynman propagator, which is defined as
\begin{align}
  &D_F(x - y) = \int \frac{\dee p}{(2\pi)^d} \frac{\ic}{p^2 - \mu_0^2 + \ic \epsilon} e^{-\ic p\cdot (x-y)} \label{eq:qft_propagator_feyn} \\
  &\qquad = \braket{0|T\phi(x)\phi(y)|0} = 
  \begin{cases}
    D(x - y) & \text{for } x^0 > y^0\\
    D(y - x) & \text{for } x^0 < y^0
  \end{cases},  
\end{align}
where $T$ denotes the time-ordered product and the relation to $D(x - y)$ can be 
found using contour integration, with the infinitesimal shift $\epsilon$ providing
a prescription for treating the poles. $D_F$ is a Green's function of the Klein-Gordon
equation
\begin{align*}
  (\Box + \mu_0^2) D_F(x-y) = -\ic \delta(x-y).
\end{align*}
The integrand of $D_F$ has poles given by the mass parameter at $p^2 = \mu_0^2$.
For an interacting theory, the poles no longer correspond to the mass parameter
$\mu_0^2$, but are shifted away from this point due to self-interaction. 
The shifted poles of the propagator then correspond to the
physical mass of a particle whereas the ``bare'' parameter $\mu_0^2$ does not.
That the poles of the propagator correspond to the particle mass 
can be seen by inserting the identity, written in terms of the (unspecified)
eigenstates of the Hamiltonian (interacting or not) and the momentum operator, 
into the expression for the propagator. The identity thus formed is
\begin{align*}
  \mathbb{I} = \ket{\Omega} \bra{\Omega} + \sum_\lambda 
               \int \frac{\dee \vec{p}}{(2\pi)^{d-1}} \frac{1}{2E(\vec{p}, \lambda)}
               \ket{\lambda_{\vec{p}}} \bra{\lambda_{\vec{p}}},
\end{align*}
where $\ket{\Omega}$ is the vacuum, $\ket{\lambda_{\vec{p}}}$ is the 
zero-momentum state $\ket{\lambda_0}$ 
boosted to momentum $\vec{p}$, and $E(\vec{p}, \lambda) = \sqrt{\vec{p}^2 + m_\lambda^2}$ 
with $m_\lambda^2$ being the mass or energy of $\ket{\lambda_0}$. 
Evaluating $\braket{\Omega | \phi(x) \mathbb{I} \phi(y) |\Omega}$ leads to the
Källén-Lehmann spectral representation of the Feynman propagator 
(see section 7 of \cite{peskin_introduction_1995})
\begin{align*}
  D_F(x - y) = \int_0^\infty \frac{\dee M^2}{2\pi} \rho_0(M^2) D_F(x - y, M^2),
\end{align*}
where $D_F(x - y, M^2)$ is the Feynman propagator with mass-parameter $M^2$
(instead of $\mu_0^2$) and
\begin{align} \label{eq:qft_spec_dens}
  \rho_p(M^2) = \sum_\lambda (2\pi) \delta(M^2 - m_\lambda^2) |\braket{\Omega|\phi(0)|\lambda_p}|^2
\end{align}
is the spectral density.
We drop the $\ket{\Omega}\bra{\Omega}$ term, since it adds at most a constant
term to the propagator.

Given that the theory contains single-particle states, the spectral density
contains a pole at $M^2 = m^2$, where $m$ is the mass of a single
particle, followed by a gap before further excitations appear. In this case, the
Feynman propagator can be separated into a one-particle contribution and the rest.
With a Fourier transform we have
\begin{align*}
  \int \dee x \, e^{\ic p.x} D_F(x - y) = \frac{\ic Z}{p^2 - m^2 + \ic \epsilon} + \dots, 
\end{align*}
where $Z$ is a real number coming from the $|\braket{\Omega|\phi(0)|\lambda_0}|^2$
factors. The single-particle term has a pole at $p^2 = m^2$, with the other
terms showing up at higher momenta.

\subsection{Interacting fields}
\label{sec:qft_interacting}

So far, we have considered the quantized free scalar field, whose solutions are
plane-waves. A free (non-interacting) field is, however, not directly relevant to 
physics, since a lack of coupling implies a lack of measurable consequences.
We can introduce interactions by adding a term to the Lagrangian density
$\mathcal{L} = \mathcal{L}_\text{free} + \mathcal{L}_\text{int}$
that leads to non-linear equations of motion. An example for real scalar field
theory, and the case we focus on in this paper, is the quartic interaction term
\begin{align*}
  \mathcal{L}_\text{int} = \frac{\lambda}{4!} \phi^4 ,
\end{align*}
where $\lambda$ is the coupling constant, or the strength of the interaction.
The resulting theory is often referred to simply as ``$\phi^4$-theory''.
Its equation of motion is
\begin{align}
  (\Box + \mu_0^2) \phi(x) = -\frac{\lambda}{3!} \phi^3,
  \label{eq:phi_4_EOM}
\end{align}
which no longer has simple plane-wave solutions. Since interacting theories
are generally difficult or impossible to solve analytically,
other approaches such as discrete (lattice-based) numerical simulation or 
perturbation theory are needed. The perturbative approach is used to study 
scattering, where it is assumed that the incoming and outgoing states far from 
the scattering location, called asymptotic states, can be described by the 
non-interacting field theory. Scattering is then represented by a unitary 
operator, the ``S-matrix'', relating the incoming and outgoing states. Elements 
of the S-matrix can be calculated perturbatively in powers of the coupling 
constant. The individual terms in the expansion have a regular form and can be 
conveniently represented using Feynman diagrams.

Since, for this work, we perform numerical simulations on a lattice, we do not go 
into perturbative calculations in detail. As mentioned above, the perturbative 
calculation of the propagator in an interacting theory reveals a shift of the 
pole mass away from the bare mass parameter $\mu_0^2$, resulting in a different,
``dressed'' physical mass $\mu^2_\text{phys}$. The mass shift is due to the 
interaction of the field with itself, which can involve modes of any momentum.
In fact, taking all possible momenta into account, the
shift diverges. Introducing a momentum cut-off into calculations, for example 
via a lattice, makes the shift dependent on this cut-off. Since the physical
mass of particles cannot diverge, and because the bare parameter is not itself
measurable, the bare mass is adjusted such that the pole of the propagator
has the correct (measured) value, even if this means that the bare mass diverges.
The procedure of adjusting bare parameters to cancel contributions from 
self-interaction is called ``renormalization''.
In general, the bare mass is not the only parameter that must be renormalized. 
Others, such as coupling constants, may also be affected.

The need for renormalization and the presence of divergent shifts can
be interpreted as signs that the theory in question is an effective 
low-energy limit of a more fundamental one \cite{peskin_introduction_1995}. 
The momentum scale where the effective theory breaks down then becomes a natural 
cut-off, such that divergent quantities are avoided. In the standard model of 
particle physics, a candidate for this cut-off is the Planck 
scale $\sim 10^{19} \,\mathrm{GeV}$, where gravitational effects are expected to 
play a significant role. However, since we don't know which theory describes
physics beyond the standard model, we also cannot know the exact location of the
cut-off, which may occur at far lower energies.

For our purposes, it is sufficient to briefly examine the only divergent
(in the absence of a cut-off)
term in $\phi^4$-theory in (1 + 1) dimensions \cite{brydges_new_1983}, 
which is the ``one-loop'' correction to the propagator. The propagator describes 
a simple ``scattering'' event involving a single incoming and outgoing particle, 
which we can examine using the same perturbative methods as are used 
for more complicated scattering events.
The Fourier-transformed propagator to first order in $\lambda$ is
\begin{align}
  \int \dee x \, &e^{\ic p.x} D_F(x - y) = \frac{\ic}{p^2 - \mu_0^2} \\ + 
    &\frac{\ic}{p^2 - \mu_0^2} 
    \underbrace{\left[
      -\ic\frac{\lambda}{2} \int \frac{\dee q}{(2\pi)^d}
      \frac{\ic}{q^2 - \mu_0^2 + \ic \epsilon}
      \right]
    }_{-\ic \delta \mu^2_1(\mu^2_0)} \frac{\ic}{p^2 - \mu_0^2} \\
  &+ \mathcal{O}(\lambda^2),
  \label{eq:qft_phi4_oneloop_amplitude}
\end{align}
where the first term is the free-field propagator \eqref{eq:qft_propagator_feyn} 
and the second term is the first-order correction. The part in square brackets 
$-\ic \delta \mu^2_1(\mu_0^2)$ diverges. 

\begin{wrapfigure}{r}{4cm}
  \includegraphics{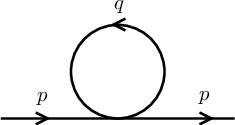}
  \caption{Feynman diagram of the one-loop correction to the free particle 
           propagator in $\phi^4$-theory.}
  \label{fig:one_loop_feynman}
  \vspace{-0.2cm}
\end{wrapfigure}
The name ``one-loop'' for the correction term comes from the corresponding 
Feynman diagram (Figure \ref{fig:one_loop_feynman}), where the two free-particle 
propagator factors outside the square
brackets correspond to incoming and outgoing particles, each with momentum 
$p$, which are represented by incoming and outgoing lines in the diagram.
The integral over $q$ inside the brackets corresponds to a ``virtual'' particle
and is represented by a loop. Additional loop-terms appear at higher orders.
Another way of writing the propagator, recognizing that the perturbative
expansion in further loop terms results in a geometric series, is as
\begin{align*}
  \int \dee x \, e^{\ic p.x} D_F(x - y) = \frac{\ic}{p^2 - \mu_0^2 - \delta \mu^2(\mu_0^2)},
\end{align*}
where the mass shift $\delta \mu^2(\mu_0^2)$ now contains all the loop-corrections. 
To first order, $\delta \mu^2(\mu_0^2) = \delta \mu^2_1(\mu_0^2)$ and the shift diverges. 
Higher-order contributions to $\delta \mu^2$ do not diverge in $(1+1)$ dimensions, such that 
removing the divergence is already achieved by adjusting the bare parameter 
$\mu_0^2$ by $\delta \mu^2_1$. A finite shift coming from higher-order 
corrections remains, but is unimportant for the purposes of investigating 
critical behavior, where the physical parameters must merely be well-defined.
There are also finite renormalization factors corresponding to the field $\phi$
and the coupling $\lambda$, which we ignore for the same reasons.

\subsection{Real scalar field theory on a lattice}
\label{sec:qft_phi4_lattice}

Discretizing the space in which a quantum field lives is a possible way of 
making field theories accessible to non-perturbative methods. It corresponds
to a dramatic reduction in the dimension of the Hilbert space and implies 
a momentum cut-off, making loop-integral contributions, which may 
be divergent in the continuum, finite on the lattice.

We use a spatial discretization to permit the use of matrix product states (MPS).
Time remains continuous, as we simulate dynamics using the Hamiltonian formalism.
For more details of this procedure, see \cite{Jordan_quantum_2011}.
We also work with an infinite lattice (in the thermodynamic limit)
since this is possible using MPS and is a more realistic setting for a field 
than a finite lattice. By assuming spatial uniformity of ground states, the 
number of variational parameters needed to approximate states remains manageable.
Our lattice version of the classical continuum theory 
introduced in \eqref{eq:phi_free_action} is given, in $(1+1)$ dimensions, 
by the Lagrangian
\begin{align*}
  L = a \sum_{n=-\infty}^\infty \left[ \frac{\dot{\phi}_n^2}{2}
      - \frac{(\phi_n - \phi_{n+1})^2}{2a^2}
      - \frac{\mu_0^2}{2} \phi_n^2 \right],
\end{align*}
where the sum is over the lattice sites, $a$ is the lattice spacing, and the 
spatial part of the derivative term has been replaced by a finite difference.
Letting $a \rightarrow 0$ recovers the Lagrangian of the continuum free scalar 
field theory. 
Applying the Euler-Lagrange equations results in
\begin{align*}
  \partial^2_t \phi_n + \frac{1}{a^2} (2\phi_n - \phi_{n-1} - \phi_{n+1}) + \mu_0^2 \phi_n = 0,
\end{align*}
where the term in brackets can be interpreted as the second derivative on the
spatial lattice. As with the classical continuum theory, a Fourier transform 
\begin{align*}
  \phi_n = \int \frac{\dee p^0}{2\pi} \int^{\pi/a}_{-\pi/a} \frac{\dee p^1}{2\pi}
           e^{-\ic p^0 x^0} e^{\ic p^1 n a} \phi(p)
\end{align*}
diagonalizes the equation of motion:
\begin{align*} 
  \int \frac{\dee p^0}{2\pi} \int^{\pi/a}_{-\pi/a} \frac{\dee p^1}{2\pi}
   e^{-\ic p^0 x^0} &e^{\ic p^1 n a} 
   \left[ \vphantom{\left(\frac{p^1 a}{2}\right)} -(p^0)^2 + \right.  \\
   &\left. \frac{4}{a^2} \sin^2 \left(\frac{p^1 a}{2}\right) + \mu_0^2 \right]
   \phi(p) = 0.
\end{align*}
The Fourier-transformed Green's function $\tilde{G}(p, a)$ satisfies 
\begin{align*}
   \left[ -(p^0)^2 + \frac{4}{a^2} \sin^2 \left(\frac{p^1 a}{2}\right) + \mu_0^2 \right]
   \tilde{G}(p, a) = -\ic,
\end{align*}
so that
\begin{align*}
  \tilde{G}(p, a) = \frac{\ic}{(p^0)^2 - \frac{4}{a^2} \sin^2 \left(\frac{p^1 a}{2}\right) - \mu_0^2},
\end{align*}
which, in the limit $a \rightarrow 0$, becomes
\begin{align*}
  \tilde{G}(p) = \frac{\ic}{p^2 - \mu_0^2},
\end{align*}
in agreement with \eqref{eq:qft_propagator_feyn}. 

Moving now to the quantized and interacting $\phi^4$-theory, we can write down the
one-loop correction to the physical mass by analogy with 
\eqref{eq:qft_phi4_oneloop_amplitude}
\begin{align*}
  &-\ic \delta \mu^2_1(\mu_0^2) = \\ 
                  &-\ic\frac{\lambda}{2} \int \frac{\dee p^0}{2\pi}
                  \int^{\pi/a}_{-\pi/a} \frac{\dee p^1}{2\pi} 
                  \frac{\ic}{(p^0)^2 - \frac{4}{a^2} \sin^2 \left(\frac{p^1 a}{2}\right) - \mu_0^2 + \ic \epsilon},
\end{align*}
which again agrees with \eqref{eq:qft_phi4_oneloop_amplitude} as $a \rightarrow 0$.
Integrating over $p^0$ using contour integration, this becomes
\begin{align*}
  -\ic \delta \mu^2_1(\mu_0^2) = -\ic\frac{\lambda}{4}
                  \int^{\pi/a}_{-\pi/a} \frac{\dee p^1}{2\pi} 
                  \frac{1}{\sqrt{\frac{4}{a^2} \sin^2 \left(\frac{p^1 a}{2}\right) + \mu_0^2}},
\end{align*}
which can be written in terms of the complete elliptic integral of the first kind
\begin{align*}
  K(k) = \int_0^{\pi/2} \dee \theta \frac{1}{\sqrt{1 - k^2 \sin^2(\theta)}}.
\end{align*}
This leaves
\begin{align}
  -\ic \delta \mu^2_1(a^2\mu_0^2) = -\ic\frac{\lambda}{2} \frac{1}{\pi} \frac{1}{\sqrt{a^2 \mu_0^2 + 4}}
                  K \left(\frac{2}{\sqrt{a^2 \mu_0^2 + 4}} \right),
  \label{eq:qft_phi4_lattice_mass_shift}
\end{align}
which is convenient for calculation using numerical computing packages, where
the elliptic integrals are commonly implemented as high-accuracy approximations.

To investigate behavior using the time-dependent variational principle,
we need the Hamiltonian form of the interacting lattice theory.
With the interaction term $\mathcal{L}_\text{int} = \frac{\lambda}{4!} \phi^4$
and using $\pi_n = \frac{\partial L}{\partial \dot{\phi_n}} =  a\dot{\phi}_n$, the 
Hamiltonian is
\begin{align*}
  H = a\sum_n \left[ \frac{\pi_n^2}{2a^2}
      + \frac{(\phi_n - \phi_{n+1})^2}{2a^2}
      + \frac{\mu_0^2}{2} \phi_n^2 + \frac{\lambda}{4!} \phi_n^4 \right].
\end{align*}
The parameters $\lambda$ and $\mu_0^2$ have dimension $[\text{mass}]^2$.
Replacing them with dimensionless quantities $\tilde{\mu}_0^2 = \mu_0^2 a^2$
and $\tilde{\lambda} = \lambda a^2$
allows us to write the dimensionless Hamiltonian $\tilde{H} = Ha$ as
\begin{align*}
  \tilde{H} = \sum_n \left[ \frac{\pi_n^2}{2}
      + \frac{(\phi_n - \phi_{n+1})^2}{2}
      + \frac{\tilde{\mu}_0^2}{2} \phi_n^2 + \frac{\tilde{\lambda}}{4!} \phi_n^4 \right],
\end{align*}
eliminating the explicit appearance of $a$. Adjusting the lattice spacing
now corresponds to altering the parameters $\tilde{\lambda}$ and $\tilde{\mu}_0^2$.
This dimensionless form is convenient for finding the continuum limit of the 
quantum critical theory (see section \ref{chap:phi4}).

Noting that $\tilde{H}$ takes the form of a many-body Hamiltonian with a
nearest-neighbor interaction, it becomes natural, especially with regard
to the later use of matrix product states, to use a basis given by 
position-space creation and annihilation operators $[a_n, a_m^\dagger] = \delta_{nm}$, 
$a_n\ket{0} = 0$ to define the quantized lattice theory. The field operator
and the conjugate momentum operator can then be defined as
\begin{align*}
  \phi_n = \frac{1}{\sqrt{2}} \left( a_n^\dagger + a_n \right) \quad \text{and} \quad
  \pi_n = \frac{\ic}{\sqrt{2}} \left( a_n^\dagger - a_n \right),
\end{align*}
such that the desired equal-time (Schrödinger picture) commutation relation
\begin{align*}
  [\phi_n, \pi_m] = \ic \delta_{nm} 
\end{align*}
is satisfied.

\subsection{Spontaneous symmetry-breaking}
\label{sec:phi4_SSB}

The $\phi^4$-theory action 
\begin{align}
  S = \int \dee x \left[ \frac{1}{2}(\partial_\mu \phi \partial^\mu \phi 
        - \mu_0^2 \phi^2) - \frac{\lambda}{4!} \phi^4 \right]
 \label{eq:phi_int_action}
\end{align}
is manifestly invariant under the discrete transformation $\phi \rightarrow -\phi$.
A given state may or may not share this symmetry. Should the ground state of a QFT
break a symmetry of the action for some set of parameters, the theory
is said to exhibit spontaneous symmetry-breaking. The word ``spontaneous'' 
refers to the fact that there are then multiple ground states (the number of
ground states is equal to the order of the symmetry), such that the actual 
ground state of the system, obtained for example by cooling, makes a seemingly
spontaneous ``choice''. 

\begin{figure}[h]
  \includegraphics[clip=true]{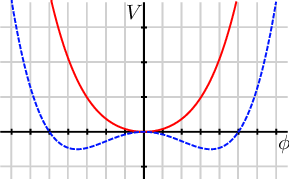}
  \caption{
    The classical effective potential in $\phi^4$-theory illustrated for
    $\mu_0^2 \ge 0$ (red) and  $\mu_0^2 < 0$ (blue dashed) showing the two possible 
    ground states for the latter case.
  }
  \label{fig:phi4_Veff_cl}
\end{figure}

Classically, the ground state lies at the minimum of a potential function.
The $\phi^4$-theory action \eqref{eq:phi_int_action} contains the 
classical effective potential
\begin{align*}
  V_\text{eff} = \frac{\mu_0^2}{2} \phi^2 + \frac{\lambda}{4!} \phi^4,
\end{align*}
which, for $\mu_0^2 \ge 0$, has a single minimum at $\phi = 0$, leaving
the symmetry $\phi \rightarrow -\phi$ intact. However, with $\mu_0^2 < 0$
there are two minima and hence two distinct ground states at 
$\pm \phi_{0, \text{cl}} > 0$ that break the symmetry, as illustrated in 
Figure \ref{fig:phi4_Veff_cl}.

The symmetry-breaking persists in the quantized theory in $(1+1)$ dimensions, which
possesses symmetric and symmetry-broken phases distinguished by the vacuum 
(ground state) expectation value of the field operator 
$\braket{\Omega | \phi | \Omega}$, henceforth abbreviated to $\braket{\phi}$.
Since the bare mass parameter $\mu_0^2$ must diverge in the continuum 
in order to renormalize the physical mass (see section \ref{sec:qft_interacting}), 
the relevant parameter is not $\mu^2_0$ as in the classical case, but the 
renormalized mass $\mu_R^2$, where we use the definition
\begin{align}
  \mu_0^2 = \mu_R^2 - \delta \mu^2_1(\mu_R^2),
  \label{eq:qft_phi4_muR}
\end{align}
with $\delta \mu^2_1(\mu_R^2)$ being the one-loop correction defined in 
\eqref{eq:qft_phi4_oneloop_amplitude},
where we perturb about the free theory with mass-squared $\mu_R^2 > 0$.
Note that $\mu_R^2$ is distinct from the physical mass ($\mu_R^2 \neq \mu_\text{phys}^2$) 
due to additional finite corrections. Since we are not generally working
under weak coupling conditions, the usual perturbative calculation of the 
full mass shift, and hence the physical mass, is not applicable, so that the
physical mass cannot be used as a parameter (although it can be found 
numerically on the lattice). 

When moving through parameter space $(\lambda, \mu_R^2)$, the transition between 
the asymmetric ground-state phase and the symmetric phase
represents a second order quantum phase transition \cite{chang_existence_1976} 
with order parameter $\braket{\phi}$. There therefore exist critical points in 
$(\lambda, \mu_R^2)$ at which the theory becomes massless and 
scale-invariant (the correlation-length $\xi$ becomes infinite). 
A scale-invariant theory should be described by dimensionless parameters, yet 
the two parameters $\lambda$ and $\mu_R^2$ have dimension $[\text{mass}]^2$ in 
$(1+1)$ dimensions. As such, the proper parameter must be the ratio 
$\lambda/\mu_R^2$. This means there is a line in parameter-space corresponding
to the critical theory.

The lattice theory also contains critical points, where the critical parameter
$\lambda/\mu_{R,c}^2 = \tilde{\lambda}/\tilde{\mu}_{R,c}^2$ now depends on the 
lattice spacing $a$ (which defines a momentum cut-off) or, equivalently, on 
$\tilde\lambda$, so that we may write $\tilde{\lambda}/\tilde{\mu}_{R,c}^2(\tilde\lambda)$. 
This dependency is 
expected to be logarithmic due to infrared corrections in the critical theory 
\cite{jackiw_how_1981} where the physical mass goes to zero. Such a dependency has 
been observed in Monte-Carlo simulations \cite{schaich_improved_2009}. To 
obtain the critical parameters of the continuum theory, we take the limit of 
$\tilde{\lambda}/\tilde{\mu}_{R,c}^2(\tilde\lambda)$ as $\tilde{\lambda} \rightarrow 0$.

Note that it is the lattice correlation length $\tilde\xi = \xi a^{-1}$ that
goes to infinity at the critical points of the lattice theory. For this reason,
there are two possible interpretations of the lattice critical point: Either as a
lattice approximation $a > 0$ to the continuum critical point where 
$\xi \rightarrow \infty$, or as a continuum limit $a \rightarrow 0$ of a
non-critical theory $\xi < \infty$. Since we are interested in the 
critical continuum theory, we will always use the former interpretation.

In the vicinity of the critical point, physical quantities scale according to  
power laws (see \cite{sachdev_quantum_2011} or section 13 of \cite{peskin_introduction_1995}). 
For the order-parameter $\braket{\phi}$, in the
symmetry-broken phase where $\braket{\phi} \neq 0$, we can thus expect
\begin{align*}
  \braket{\phi} = A(\tilde{\lambda})\left[ 
                  \frac{\tilde{\lambda}}{\tilde{\mu}_R^2}
                  - \frac{\tilde{\lambda}}{\tilde{\mu}_{R,c}^2}  
                  \right]^{\beta(\tilde{\lambda})},
\end{align*}
where $A(\tilde{\lambda})$ is some constant and $\beta(\tilde{\lambda})$
is the critical exponent. We also define the scaling for the energy (or mass) of the
lowest-lying excitation:
\begin{align*}
  \Delta E = B(\tilde{\lambda})\left| 
                  \frac{\tilde{\lambda}}{\tilde{\mu}_R^2}
                  - \frac{\tilde{\lambda}}{\tilde{\mu}_{R,c}^2}  
                  \right|^{\nu(\tilde{\lambda})}.
\end{align*}
The energy $\Delta E$ should correspond to the particle mass $\mu_\text{phys}$
(given by poles in the propagator) in the symmetric phase, but may belong to
a topologically non-trivial soliton (kink) excitation in the symmetry-broken 
phase (providing a localized transition $\phi \rightarrow -\phi$ between
two different ground states at $\vec{x} \rightarrow \pm \infty$).

Predictions for the critical exponents $\beta$ and $\nu$ in the limit 
$\tilde{\lambda} \rightarrow 0$ can be obtained based on the 
universality principle, which comes from renormalization group theory 
\cite{sachdev_quantum_2011}.
$\phi^4$-theory in $(1+1)$ dimensions has been shown to be a continuum
limit of the transverse Ising model \cite{simon_phi4_2_1973} and, as such,
is predicted to share its critical exponents. For $\braket{\phi}$
we thus expect $\beta = 1/8$ and for $\Delta E$ we expect
$\nu = 1$. For more information on the critical behavior of $\phi^4$-theory,
including a derivation of these critical exponents, see \cite{kleinert_critical_2001}.
The critical parameter $\lambda/\mu_{R,c}^2$ is not a universal quantity,
depending instead on the particulars of $\phi^4$-theory. It is also not 
accessible to perturbative techniques \cite{chang_existence_1976}, making it an 
interesting target for lattice methods. We estimate it, as well as the critical 
exponents defined above, in section \ref{chap:phi4}.

\section{Matrix product states}
\label{chap:MPS}

\label{sec:mps_intro}

Matrix product states (MPS) are pure states of one-dimensional lattice systems
with a particular form that puts a limit on the amount of entanglement a state
can contain. The amount of entanglement is related to the bond-dimension $D$,
which is the dimension of the matrices that make up the state coefficients.
Most quantities (assuming open boundary conditions) can be calculated with 
complexity $\mathcal{O}(ND^3)$, where $N$ is the number of lattice sites or, in
the case of uniform (translation invariant) MPS in the thermodynamic limit (uMPS), 
the number of necessary solver iterations.

In this section, we define uMPS and derive an implementation of 
the time-dependent variational principle (TDVP) as well as a related ansatz for 
determining excitation energies. Both algorithms were first described by Haegeman et al.
\cite{haegeman_time-dependent_2011, haegeman_variational_2012}. We also set
out a variational conjugate-gradient algorithm, demonstrating significantly
improved convergence speeds for $\phi^4$-theory compared to the TDVP.

\subsection{Uniform MPS in the thermodynamic limit}
\label{sec:mps_umps}

Uniform matrix product states (uMPS) have the form
\begin{align*}
\ket{\Psi(A)} = \sum_{\{s\}=0}^{d-1}
                v_L^\dagger 
                \left[\prod_{i=-\infty}^{+\infty} A^{s_i} \right] 
                v_R
			    \ket{\vec{s}},
\end{align*}
where $\ket{\vec{s}} = \ket{\dots s_1 \dots s_N \dots}$ and 
the site-independent $d \times D \times D$ tensor $A$ contains the 
parameters for the entire state.
The boundary vectors $v_L$ and $v_R$ are of length $D$ and are irrelevant in
calculations so that we may ignore them. The uMPS states form a sub-manifold
of Hilbert space $\mathcal{M}_\text{uMPS} \subset \mathcal{H}$ that depends on
$D$. They have $dD^2$ complex parameters, but possess one non-physical degree of
freedom corresponding to the norm and $D^2 - 1$ gauge degrees of freedom due
to invariance under transformations
\begin{align} \label{eq:umps_gauge_transf}
  A^s &\rightarrow g A^s g^{-1},
\end{align}
where the trivial transformation $g = c\mathbb{I}$ is not counted,
so that the number of physical degrees of freedom is 
$(dD^2 - 1) - (D^2 - 1) = D^2(d-1)$.
The norm is determined by the infinite power of the $D^2 \times D^2$ matrix 
$E = \sum_s A^s \otimes \overline{A^s}$ so that the spectral radius of 
$E$ must be one: $\rho(E) = 1$. 
We further require that $E$ has a unique eigenvalue of greatest magnitude, 
which must then be equal to one in order to obtain well-defined expectation
values and to avoid dependencies on the boundary vectors \cite{haegeman_geometry_2012}.

The left and right eigenvectors of $E$ with eigenvalue 1 we name $\bra{l}$ and $\ket{r}$ 
respectively. Via the Choi-Jamiolkowsky isomorphism, we may also
define $D \times D$ matrices $l$ and $r$ such that $\sum_{s} {A^s}^\dagger l A^s = l$
and $\sum_{s} A^s r {A^s}^\dagger = r$. Numerical computation of quantities
involving $E$ is more efficient in this matrix representation, scaling with $\mathcal{O}(D^3)$
rather than $\mathcal{O}(D^6)$ (assuming a naive matrix-multiplication algorithm).
Single-site expectation values can be computed as
\begin{align*}
  \braket{\Psi(A)|o|\Psi(A)} = \braket{l|E^o|r} 
  = \tr\left[l \sum_{s,t} A^t r {A^s}^\dagger \braket{s|o|t} \right],
\end{align*}
where $E^o = \sum_{s,t} \braket{s|o|t} A^t \otimes \overline{A^s}$.

The gauge freedom \eqref{eq:umps_gauge_transf} implies that there is no unique MPS 
representation of a given state. There are, however, useful forms for the uMPS tensor $A$, of which 
the so-called right canonical form has the properties
\begin{align}
  \label{eq:umps_rcf}
  \sum_{s} A^{s} {A^{s}}^\dagger = \mathbb{I}_D \iff r = \mathbb{I}_D \quad 
  \text{and} \\
  l_{\alpha\beta} = \delta_{\alpha\beta} \lambda_\alpha^2
  \quad (\alpha,\beta = 1\dots D),
\end{align}
where $\lambda_\alpha$ are the Schmidt coefficients corresponding
to decomposing the system into two infinite halves 
$\ket{\Psi} = \sum_{\alpha=1}^D \lambda_\alpha \ket{\psi_L} \otimes \ket{\psi_R}$,
with orthonormal Schmidt vectors for the left and right halves $\ket{\psi_L}$ and
$\ket{\psi_R}$. 
This makes explicit the relationship
between the amount of entanglement possessed by a uMPS state
and the bond dimension $D$, which is equal to the
Schmidt rank of the half-chain decomposition. Since the Schmidt coefficients
are also the eigenvalues of the density matrix corresponding to the reduced
state on the half-chain, we can easily calculate the corresponding
von Neumann entropy as
\begin{align} \label{eq:umps_S_hc}
  S = -\sum_{\alpha=1}^D \lambda_\alpha^2 \log_2 \lambda_\alpha^2.
\end{align}
The conditions \eqref{eq:umps_rcf} fix all gauge degrees of freedom and
$A$ can always be made to fulfill them by performing a gauge-transformation. 
This can be verified using the eigenvalue
equations $E\ket{r} = \ket{r}$ and $\bra{l}E = \bra{l}$:
We find that a gauge transformation affects $l$ and $r$ as
$l \rightarrow g^{-1\dagger} l g^{-1}$ and $r \rightarrow g r g^\dagger$
which, together with \eqref{eq:umps_rcf}, fully specify the $g$ needed to
put an arbitrary $A$ into canonical form.

To implement the TDVP for uMPS, we need to understand the tangent plane 
$\mathbb{T}_{\ket{\Psi(A)}}$ to $\mathcal{M}_\text{uMPS}$ at a point $\ket{\Psi(A)}$. 
Uniform tangent vectors have the form
\begin{align}
  &\ket{\Phi(B)} = \sum_{i=1}^{d D^2} B_{i} \ket{\partial_{i}\Psi(A)} 
  \label{eq:mps_uni_tangvec} \\
  & \; = \sum_{n=-\infty}^{+\infty} \sum_{\{s\}=0}^{d-1} v_L^\dagger 
    \left[\prod_{i=-\infty}^{n-1} A^{s_i} \right]
    B^{s_n}
    \left[\prod_{i=n+1}^{+\infty} A^{s_i} \right]
    v_R \ket{\vec{s}},
\end{align} 
where we use the shorthand notation 
$\ket{\partial_{i}\Psi(A)} \equiv \partial/\partial A_{i} \ket{\Psi(A)}$
with the index $i$ running over all entries of the tensors $A$ and $B$.
We call the tensor $B$ the parameter-space tangent vector. Changing
the state parameters as $A \rightarrow A + \dee \tau B$ changes
the state as $\ket{\Psi(A)} \rightarrow \ket{\Psi(A)} + \dee \tau \ket{\Phi(B)}$.
We also define ``boosted'' tangent vectors for uniform systems
\begin{align}
  &\ket{\Phi_p(B)} = \label{eq:mps_uni_tangvec_p} \\
  & \, \sum_{n=-\infty}^{+\infty} e^{\ic pn} \sum_{\{s\}=0}^{d-1} v_L^\dagger 
    \left[\prod_{i=-\infty}^{n-1} A^{s_i} \right]
    B^{s_n}
    \left[\prod_{i=n+1}^{+\infty} A^{s_i} \right]
    v_R \ket{\vec{s}},
\end{align} 
representing different momentum sectors $p$
such that $\ket{\Phi_0(B)} \equiv \ket{\Phi(B)}$.
These are useful for studying excitations
(see section \ref{sec:mps_uniform_excitations}).

As with the state $\ket{\Psi(A)}$, there are non-physical degrees of freedom
in the parameter tensor $B$.
Apart from the state itself lying in the tangent place $\ket{\Psi(A)} \in \mathbb{T}_{\ket{\Psi(A)}}$, 
they also include infinitesimal gauge transformations such that a tangent vector is
invariant under $\ket{\Phi_p(B)} \rightarrow \ket{\Phi_p(B + \mathcal{N}_p(x))}$
with $\mathcal{N}^s_p(x) = e^{-\ic p}xA^s - A^sx$. That $\ket{\Phi_p(\mathcal{N}_p(x))}$ 
corresponds to an infinitesimal gauge transformation can be checked by using
one-parameter site-dependent gauge transformation matrices 
$g_n(\eta) = \mathbb{I} + \eta x e^{\ic pn}$ to transform the state ($A^s_n \rightarrow g_{n-1} A^s_n g_n^{-1}$),
taking the derivative $\left. \dee/\dee \eta \ket{\Psi(A)} \right|_{\eta = 0}$
to obtain the infinitesimally transformed state.

All non-physical degrees of freedom can be eliminated by requiring
that $B$ satisfy a gauge-fixing condition such as
the right gauge-fixing condition
\begin{align}
  \sum_{s} B^{s} r {A^{s}}^\dagger = 0 = E^{B}_{A} \ket{r}.
  \label{eq:mps_uni_right_gfc}
\end{align}
Note that $\braket{\Phi_p(B)|\Psi(A)} = 2\pi \delta(p) \braket{l|E^B_A|r}$
where the second factor is zero according to \eqref{eq:mps_uni_right_gfc}. 
Hence, this condition also includes orthogonality to the ground state 
$\ket{\Psi(A)}$ for momentum zero, which cannot be obtained by a mere 
gauge transformation. If we start with an arbitrary B, then for momentum zero 
we have to manually impose $\braket{l|E^B_A|r} = 0$ (orthogonality to the
ground state), after which we can bring it into a form where it satisfies 
\eqref{eq:mps_uni_right_gfc} by doing a gauge transformation. 
For non-zero momentum, a gauge transformation alone is sufficient.
We can see this by making the replacement $B \rightarrow B + \mathcal{N}_p(x)$ 
in \eqref{eq:mps_uni_right_gfc}, resulting in
\begin{align*}
  \ket{xr} = (E - \mathbb{I} e^{-\ic p})^{-1}E^B_A\ket{r},
\end{align*}
which we can solve to obtain $x$. For $p \neq 0$, the solution is unique 
(assuming $r$ is full-rank). In case $p=0$ the inverse must 
become a pseudo-inverse, leaving freedom $x \rightarrow x + c\mathbb{I}$
corresponding to the null space of $(E - \mathbb{I})$, which we eliminated
from $E^B_A\ket{r}$ by imposing orthogonality to the ground state. 
However, for $p = 0$ this freedom in $x$ is not 
part of the gauge group: $\mathcal{N}_0(c\mathbb{I}) = 0$ so that the condition
fixes exactly the gauge (and norm for $p=0$) degrees of freedom.
Restricting $B$ so that it always satisfies \eqref{eq:mps_uni_right_gfc} can
be achieved using the parametrization
\begin{align}
  B^s(x) = l^{-1/2} x V^s r^{-1/2},
  \label{eq:mps_tdvp_uni_B_param}
\end{align}
where $x \in M_{D \times D(d - 1)}$ and the
$D(d - 1) \times dD$  matrix 
$[V]_{(\alpha,s);\beta} = [V^s]_{\alpha\beta}$
(where the index $(\alpha, s)$ combines $s$ and $\alpha$)
is defined so that $V^\dagger$ contains an orthonormal basis 
($VV^\dagger = \mathbb{I}$) for the null-space of $R^\dagger$, with
\begin{align*}
  [R]_{(\alpha, s); \beta} = [r^{1/2} {A^s}^\dagger]_{\alpha, \beta},
\end{align*}
resulting in $VR = 0$.

\begin{figure}[t]
  \fbox{
      \begin{minipage}{0.45\textwidth}
        \textbf{The time-dependent variational principle} \\
        \justifying
        \begin{wrapfigure}{r}{4.0cm}
          \vspace{-10pt}
          \hspace{-20pt}
          \includegraphics{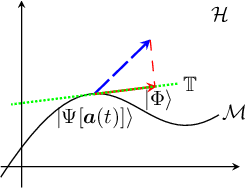}
          \vspace{-20pt}
        \end{wrapfigure}
        We wish to compute the time evolution of a quantum state.
        Because the dimension of the Hilbert space is large, 
        we restrict ourselves to a class of relevant states $\ket{\Psi[\vec{a}]}$
        with a manageable number of parameters $\vec{a} \in \mathbb{C}^d$, $d \ll \dim(\mathcal{H})$.
        This defines a sub-manifold $\mathcal{M} \subset \mathcal{H}$.
        Given a starting state $\ket{\Psi[\vec{a}(t)]}$, the Schrödinger equation
        gives us the infinitesimal evolution
        $\ket{\Psi(t + \dee t)} = \ket{\Psi[\vec{a}(t)]} - \mathrm{i} \dee t H \ket{\Psi[\vec{a}(t)]}$,
        where generally $\ket{\Psi(t + \dee t)} \notin \mathcal{M}$ because the step
        $-\mathrm{i}H \ket{\Psi[\vec{a}(t)]}$ (blue dashed arrow) 
        need not lie within the tangent plane $\mathbb{T}$ to $\mathcal{M}$.
        To optimally approximate time evolution whilst remaining
        in $\mathcal{M}$, we project the exact step onto $\mathbb{T}$,
        which means finding a tangent vector $\ket{\Phi} \in \mathbb{T}$
        (solid red arrow) that minimizes
        $\| \mathrm{i}H\ket{\Psi[\vec{a}(t)]} + \ket{\Phi}\|^2$.
        A tangent vector has the form $\ket{\Phi[\vec{b}]} = b^j \ket{\partial_j \Psi}$
        (with $\ket{\partial_j \Psi} = \partial/\partial a^j \ket{\Psi[\vec{a}]}$),
        leading to the flow equations
        \begin{align*}
            \ic \dot{a}^j(t) = g^{jk}&\braket{\partial_k
                                      \Psi| H | \Psi}
            \iff \\
            &\ic \ket{\Phi[\dot{\vec{a}}(t)]} = \ket{\partial_j \Psi} 
                                               g^{jk}\braket{\partial_k
                                               \Psi| H | \Psi},
        \end{align*}    
        where $g^{jk}$ is the inverse of the pullback metric
        $g_{jk} = \braket{\partial_j \Psi| \partial_k \Psi}$
        (assuming $g_{jk}$ has no kernel). We identify 
        $\ket{\partial_j \Psi} g^{jk}\bra{\partial_k \Psi}$
        as the projector onto $\mathbb{T}$.
        As a simplification, we have taken
        $\ket{\Psi(\vec{a})}$ to be always normalized and to be 
        a holomorphic function of $\vec{a}$.
      \end{minipage}
  }
  \caption{}
  \label{fig:TDVPbox}
\end{figure}

\subsection{Time-dependent variational principle for matrix product states}
\label{sec:mps_tdvp}
\label{sec_mps_umps_tdvp}

To apply the time-dependent variational principle (TDVP | see Figure \ref{fig:TDVPbox}) 
to uMPS we have
to find an $x$ that satisfies
\begin{align*}
  x = \arg \min_{x'} \left|\left| \ket{\Phi(B(x'))} + \ic H \ket{\Psi(A)} \right|\right|
\end{align*}
for a given $A$, where $B(x)$ is the gauge-fixing parametrization 
\eqref{eq:mps_tdvp_uni_B_param} and $\ket{\Phi(B)}$ is a uniform tangent vector
as defined in \eqref{eq:mps_uni_tangvec}.
We minimize the expression by setting its derivative with respect to $x^\dagger$
equal to zero. To do this, we need to calculate the two terms containing $x^\dagger$.
The first is the tangent vector norm
\begin{align}
  \eta \equiv \braket{\Phi(B)|\Phi(B)} \equiv \overline{B^i} B^j g_{ij},
  \label{eq:mps_tdvp_term1}
\end{align}
where the summation indices $i$ and $j$ run over all the entries of $B$.
With the gauge-fixing parametrization, this simplifies to
\begin{align} \label{eq:mps_tdvp_eta}
  \eta(x) \equiv \braket{\Phi(B(x))|\Phi(B(x))} 
    = |\mathbb{Z}|\tr\left[x^\dagger x \right], 
\end{align}
where $|\mathbb{Z}|$ represents the size of the infinite lattice.
We also have the Hamiltonian term
\begin{align}
  \braket{\Phi(B)|H-\braket{H}|\Psi(A)} \equiv \overline{B^i} \braket{\partial_{i} \Psi |H-\braket{H}|\Psi(A)},
  \label{eq:mps_tdvp_term2}
\end{align}
where we are free to subtract $\braket{H} = \braket{\Psi(A)|H|\Psi(A)}$
without changing the result due to \eqref{eq:mps_uni_right_gfc}, which
ensures $\braket{\Phi(B)|\Psi(A)} = 0$.
Assuming the Hamiltonian is uniform and can be written as a 
sum of nearest-neighbor terms $H = \sum_n h_{n,n+1}$, this
simplifies to
\begin{align*}
  \braket{\Phi(B(x))|H-\braket{H}|\Psi(A)}
    = |\mathbb{Z}| \tr\left[x^\dagger F \right],
\end{align*}
with
\begin{align*}
  F &= \sum_{s} l^{1/2} A^s K r^{-1/2} {V^s}^\dagger \\
  &+ \sum_{s,t} l^{1/2} C^{s,t} r {A^t}^\dagger r^{-1/2} {V^s}^\dagger \\
  &+ \sum_{s,t} l^{-1/2} {A^t}^\dagger l C^{t,s} r^{1/2} {V^s}^\dagger.
\end{align*}
$K$ contains the sum of Hamiltonian terms over one half of the infinite lattice
\begin{align*}
  \ket{K} = \sum_{n=0}^{+\infty} (E)^n E^C_{AA} \ket{r},
\end{align*}
with 
\begin{align}
  C^{s,t} = \sum_{u,v} \braket{s,t|h - \braket{h}|u,v} A^u A^v
  \label{eq:umps_tdvp_gen_C}
\end{align}
so that
\begin{align}
  E^{C}_{AB} = \sum_{s,t} C^{s,t} \otimes &\overline{A^s B^t}
  \iff \\
  &E^{C}_{AB}\ket{x} \approx \sum_{s,t} C^{s,t} x {B^t}^\dagger {A^s}^\dagger
  \label{eq:umps_tdvp_gen_EC}
\end{align}
represents a single term in $H - \braket{H}$ acting on a pair of sites.
Since $E$ has a unique eigenvalue of largest magnitude with value $1$, 
we can split such infinite sums into two parts
\begin{align*}
  \ket{K} = 
            \sum_{n=0}^{+\infty} Q(QEQ)^nQ E^C_{AA} \ket{r} 
            + |\mathbb{N}|\ket{r}\braket{l|E^C_{AA}|r},
\end{align*}
with the projector $Q = Q^n = \mathbb{I} - \ket{r}\bra{l}$ leading to 
$\rho(QEQ) < 1$, turning the first term into a geometric series
\begin{align*}
   \sum_{n=0}^{+\infty} Q(QEQ)^nQ = Q(\mathbb{I} - QEQ)^{-1}Q,
\end{align*}
whilst the second term is zero due to 
$\braket{l|E^C_{AA}|r} = \braket{\Psi(A)|h - \braket{h}|\Psi(A)} = 0$. 
We thus have
\begin{align*}
  \ket{K} = Q(\mathbb{I} - QEQ)^{-1}Q E^C_{AA} \ket{r} = (\mathbb{I} - E)^\text{P} E^C_{AA} \ket{r},
\end{align*}
where $\text{P}$ denotes the pseudo-inverse. $\ket{K}$ can be calculated directly, but
would involve $\mathcal{O}(D^6)$ operations. Instead, we avoid the inverse
by re-arranging to give
\begin{align}
  (\mathbb{I} - QEQ) \ket{K} = Q E^C_{AA} \ket{r},
  \label{eq:mps_tdvp_uni_K_mat}
\end{align}
which can be solved in the matrix representation for $K$ with complexity 
$\mathcal{O}(D^3)$ using a sparse solver. 

Finally, we obtain the TDVP flow equations
\begin{align*}
  \dot{A}^s = -\ic B^s(F),
\end{align*}
giving us the time-evolution of $\ket{\Psi(A)} \in \mathcal{M}_\text{uMPS}$ that
best approximates the exact (Schrödinger) evolution.
We may integrate them numerically using the Euler method with the following
algorithm:
\begin{enumerate}
  \item Calculate $F$ (including prerequisites $C$, $K$).
  \item Take a step by setting $A^s(t + \dee t) = A^s(t) - \ic \dee t B^s(F)$.
  \item Restore canonical form of $A$ using a gauge transformation.
  \item Compute $l$ and $r$ and normalize, then compute other 
        desired quantities, such as the energy, and adjust the step size 
        $\dee t$ as required.
\end{enumerate}
Normalization is necessary despite gauge-fixing because we take finite 
time steps along tangent vectors. For the same reason, the gauge degrees
of freedom will also drift so that we must perform a gauge transformation
if we wish to maintain canonical form (which reduces computational 
requirements due to the simple forms of $l$ and $r$).
Determining the eigenvectors $\bra{l}$ and $\ket{r}$ can be done 
iteratively (using a sparse eigensolver) 
with per-iteration complexity $\mathcal{O}(D^3)$. 
If the corresponding eigenvalue is not $1$ then $A$ 
should be scaled appropriately to normalize the state.
The total complexity of the algorithm is $\mathcal{O}(n_\text{itr} D^3)$,
where $n_\text{itr}$ is the number of iterations required to find $\bra{l}$ and $\ket{r}$
plus the solver iterations needed to obtain $K$ using 
\eqref{eq:mps_tdvp_uni_K_mat}.

\subsubsection{Imaginary time evolution}
\label{sec_tdvp_imtime}

Imaginary-time evolution can be seen as a gradient-following minimization
method applied to the energy functional $H(\overline{\Psi}, \Psi) = \braket{\Psi|H|\Psi}$. Taking the 
first derivative with respect to $\bra{\Psi}$ results in 
$\frac{\dee H(\Psi)}{\dee \bra{\Psi}} = H \ket{\Psi}$ so that a small step $\dee \tau$ 
in the direction $-H \ket{\Psi}$ should take us closer to the ground state
(given that $\ket{\Psi}$ is not orthogonal to it).
The same result is obtained by replacing $t$ with $-\ic \tau$ in the
Schrödinger equation, hence ``imaginary-time evolution''.

The TDVP flow equations can be used to efficiently approximate the exact imaginary-time
evolution by making the same replacement $t \rightarrow -\ic \tau$. 
If we start with a state in some variational class that is not orthogonal
to the exact ground state, integrating the flow equations will then locate the
best ground-state approximation within the class unless we get stuck in a local
minimum. The norm $\eta$ (defined in \eqref{eq:mps_tdvp_eta}) 
of the approximate evolution vector $\ket{\Phi}$
 acts as a convergence measure: 
It represents the size of
the gradient $H\ket{\Psi}$ as projected onto $\mathbb{T}$, which
goes to zero at the energetic minimum. However, it also goes to zero
at local minima of $\braket{\Psi(\vec{a})|H|\Psi(\vec{a})}$
so that some caution must be used in interpreting it.

Note that, unlike with real-time evolution, any errors made in integrating
the imaginary-time flow equations do not accumulate because an
accurate step will always take the state closer to the ground state 
irrespective of previous steps. The convergence of the energy expectation
value is quadratic in $\eta$
\begin{align*}
  \frac{\dee}{\dee \tau} \braket{H} = -2 \eta^2
\end{align*}
so that the approximate ground state energy can be obtained, to a given 
precision, with less effort than the ground-state expectation value of 
a general observable.

\subsubsection{Conjugate gradient algorithm for finding ground states}
\label{sec:mps_CG}

There are a wide range of unconstrained minimization algorithms available
that often provide far better convergence than simply taking finite steps along the 
gradient, including the non-linear conjugate-gradient (CG) method for approximately
quadratic functions (see appendix \ref{sec:min_CG}). Applying such techniques
to quantum states restricted to a variational manifold may allow us to find 
ground states more efficiently than by integrating the imaginary-time 
TDVP flow equations. Here we present a naive variational implementation of the 
non-linear CG method that can be implemented using only the tools already needed 
for the TDVP.
Together with gauge-fixing conditions, it is well-defined for uMPS. For more information
about the differential-geometric properties of $\mathcal{M}_\text{uMPS}$, see
\cite{haegeman_geometry_2012}.
For more details about optimization
on Riemannian manifolds, see \cite{absil_optimization_2009}. 

The function to minimize is 
$H(\overline{\vec{x}}, \vec{x}) = \braket{\Psi(\vec{x})|H|\Psi(\vec{x})}$,
which is approximately quadratic in the variational parameters $\vec{x}$ 
near any stationary points. The key difference to the standard CG method is 
the introduction of the non-trivial
parameter metric 
$g_{ij}(\vec{x}) = \braket{\partial_{i} \Psi(\vec{x})|\partial_j \Psi(\vec{x})}$.
For each step $n$ of the algorithm, we require 
the gradient with respect to $\vec{x}_n$, which is given by 
$r_n^j = g^{ij} \braket{\partial_{i} \Psi(\vec{x}_n) | H |\Psi(\vec{x}_n)}$
and which we can calculate by minimizing 
$||\ket{\Phi(\vec{r}_n)} + H \ket{\Psi(\vec{x}_n)}||$, as with the TDVP.
We also need the factor
\begin{align*}
  \beta_n = \frac{\vec{r}_{n+1} . \vec{r}_{n+1}
                 }{
                  \vec{r}_n.\vec{r}_n
                 }
          = \frac{\overline{\vec{r}}_{n+1}^{{i}} \vec{r}_{n+1}^j g_{{i}j}
                 }{
                  \overline{\vec{r}}_{n}^{{i}} \vec{r}_{n}^j g_{{i}j}
                 },
\end{align*}
which we can again calculate using methods already needed for the TDVP.

Additional work is required, however, because 
each iteration of the algorithm involves making a step of length $\alpha$ 
that minimizes $H$ along a given direction $\vec{p}_i$. To do this in curved space, 
we should follow a geodesic. 
Also, to obtain $\vec{p}_i$
we must add tangent vectors $\vec{r}_i$ and $\beta_{i-1} \vec{p}_{i-1}$
belonging to tangent planes at different points
$\vec{x}_i$ and $\vec{x}_{i-1}$, requiring the parallel transport of $\vec{p}_{i-1}$.
This adds significantly to the complexity of the algorithm.
However, if $g(\vec{x})$ is well-behaved such that the parallel-transport 
map is approximately trivial then we can make steps using
$\vec{x}_{n+1} \approx \vec{x}_n + \alpha_n \vec{p}_n$
with $\vec{p}_{n} \approx \vec{r}_n + \beta_{n-1} \vec{p}_{n-1}$.
Whether this assumption is reasonable depends on the particular combination of 
system and variational class. Nevertheless, should it not hold, the 
line-search used to find the step-size still guarantees that the energy will 
fall with each step, so that failure is not catastrophic and merely leads to 
slower convergence.

\begin{figure}[h]
    \includegraphics{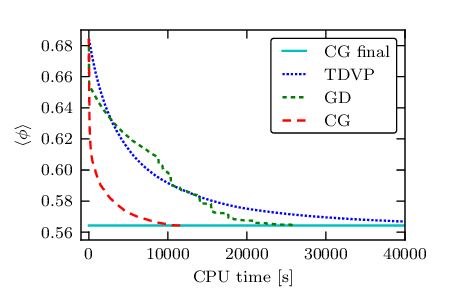}
    \vspace{-0.5cm}
  \caption{
    Convergence of the field expectation value $\braket{\phi}$ with CPU time for the 
    conjugate gradient (CG) method versus imaginary-time evolution via Euler
    integration of the TDVP flow equations and gradient descent 
    (GD | stepping along the gradient as with the TDVP, but using a 
    line-search to determine the size of each step by minimizing the energy). 
    The model is $\phi^4$-theory (as defined 
    in section \ref{sec:qft_phi4_lattice}) with parameters $\tilde{\lambda} = 0.2$ and 
    $\tilde{\lambda} / \tilde{\mu}_R^2 = 69$. The bond-dimension is $D = 64$ and
    the stopping criterion is $\eta < 10^{-6}$.
    The line (``CG final'') indicates the final value taken from the CG curve. 
    We use the same line-search algorithm for both the CG and GD methods.
    The discontinuities in the GD curve are large jumps that could occasionally be made in a
    particular direction.
  }
  \label{fig:CG_vs_brute_example}
\end{figure}

In this work, we observe that the above naive method is 
highly effective in the case of uMPS applied to lattice
$\phi^4$-theory near its critical point, as exemplified in
Figure \ref{fig:CG_vs_brute_example}. To further improve efficiency,
we also implement some additional optimizations:
For near-critical systems, the slowest part of the algorithm, which is also the 
bottleneck for the
TDVP algorithm, is the determination of the eigenvectors $l$ and $r$ of $E$,
which we do iteratively. When taking small 
(imaginary) time steps, convergence speed improves when using $l$ and $r$ from 
a nearby state (e.g. the previous step) as a starting point for the iteration. 
In the CG algorithm, each evaluation of $H$ for some $\alpha$ visited during the 
line-search requires $l$ and $r$ to be determined. To speed this up we store $l$ and $r$ 
for each point visited, using the closest (in terms of $\alpha$) stored copies 
as starting points for the iteration at each new point visited. Also, we 
do not demand the optimal value of $\alpha$ to high precision, since
conjugacy will eventually be lost anyway due to the assumptions made and because
the target function is not exactly quadratic. This usually reduces the number of 
evaluations of $H(\overline{A}, A)$ to less than ten for each CG iteration. We use the same
optimized line-search routine to determine the step size for the gradient-descent
results in Figure \ref{fig:CG_vs_brute_example}.

We also observe improved convergence of the CG method when performing a small
number of TDVP steps (of fixed step-size) after each reset of the CG algorithm.

\subsection{Excitations with uniform matrix product states}
\label{sec:mps_uniform_excitations}

Given a set of trial
states $\ket{\Phi(\vec{b})}$ linear in their parameters and orthogonal to the
ground state, the stationary points of the energy functional 
$H(\overline{\vec{b}}, \vec{b}) = \braket{\Phi(\vec{b})|H|\Phi(\vec{b})} / 
\braket{\Phi(\vec{b})|\Phi(\vec{b})}$
represent approximate excited states.
These can be found by solving the generalized eigenvalue equation
\begin{align}
  \mathbf{H} \vec{b} = E \mathbf{N} \vec{b},
  \label{eq:tivp_rayleigh_ritz}
\end{align}
where $\overline b^{s} \mathbf{H}_{st} {b'}^t = \braket{\Phi(\vec{b})|H|\Phi(\vec{b}')}$
and $\overline b^{s} \mathbf{N}_{st} {b'}^t = \braket{\Phi(\vec{b})|\Phi(\vec{b}')}$.
Given a uMPS approximate ground state, the uMPS boosted tangent plane 
\eqref{eq:mps_uni_tangvec_p} represents a good set of ansatz states for 
probing low-lying excitations of uniform systems \cite{haegeman_variational_2012}
using this method.

The suitability of the tangent vectors as ansatz-states is based on the ideas of 
Bijl, Feynman and Cohen and assumes that elementary excitations are momentum 
superpositions of local disturbances of the ground state. Where there is
more than one ground state, such as in the case of spontaneous 
symmetry-breaking, elementary excitations may also involve their combination to 
form topologically non-trivial states (for example, kink solutions).
For this reason, we additionally include the case where a local disturbance 
interpolates between two degenerate ground states. We write the resulting states
as
\begin{align}
  &\ket{\Phi_p(B;A,\tilde{A})} = \label{eq:mps_excite_ansatz} \\ 
                    & \quad \sum_{n \in \mathbb{Z}} e^{\ic pn}
                    \sum_{\{s\}=0}^{d - 1}
                    v_L^\dagger \left[\prod_{i=-\infty}^{n-1} A^{s_i} \right]
                    B^{s_n} \left[\prod_{i=n+1}^{+\infty} \tilde{A}^{s_i} \right]
                    v_R \ket{\vec{s}},
\end{align}
where $A = \tilde{A}$ recovers the boosted tangent vectors for uMPS
\eqref{eq:mps_uni_tangvec_p} and setting $A$ and $\tilde{A}$ to be the uMPS
parameters for two different ground states gives us topologically non-trivial
excitations. With these ansatz states, which are linear in the parameters $B^s$,
excitation energies can be obtained by solving 
\eqref{eq:tivp_rayleigh_ritz}, which in this case becomes
\begin{align*}
  \mathbf{H}_p B_i = \Delta E_i \mathbf{N}_p B_i,
\end{align*} 
where the index $i$ denotes the $i$th solution, $B_i$ is a vector of length 
$dD^2$ containing the entries of each $B^s$ and 
the matrices $\mathbf{H}_p$ and $\mathbf{N}_p$ are defined as
\begin{align*}
  2\pi\delta(p' - p) B^\dagger \mathbf{H}_p B' &= \braket{\Phi_{p'}(B)|H - \braket{H}|\Phi_p(B')} \quad \text{and}\\
  2\pi\delta(p' - p) B^\dagger \mathbf{N}_p B' &= \braket{\Phi_{p'}(B)|\Phi_p(B')},
\end{align*}
where we subtract the ground-state energy 
$\braket{H} \equiv \braket{\Psi(A)|H|\Psi(A)}$ 
so as to obtain a finite eigenvalue $\Delta E_i$, 
which is thus the energy difference between the excited state and the 
ground state.

The effective Hamiltonian term $B^\dagger \mathbf{H}_p B'$ contains three
(infinite) sums over the lattice sites: One from each $\ket{\Phi_p(B)}$
and one from the Hamiltonian. Terms where $B$ and $B'$
occur at different lattice sites $n$ and $n' \neq n$ acquire a factor 
$e^{\ic p (n-n')}$. Infinite sums occur over powers of $E^A_A$,
$E^{\tilde{A}}_{\tilde{A}}$, $e^{+\ic p} E^A_{\tilde{A}}$ and 
$e^{-\ic p} E^{\tilde{A}}_A$, where the first 
two have spectral radius $1$ and can be calculated with techniques used in the
TDVP algorithm (section \ref{sec_mps_umps_tdvp}), leading to pseudo-inverse factors $(\mathbb{I} - E^A_A)^\mathrm{P}$. 
$E^A_{\tilde{A}}$ and $E^{\tilde{A}}_A$ are related to the overlap between the 
two ground states. The per-site fidelity is equal to the spectral radius 
$\rho(E^A_{\tilde{A}}) = \rho(E^{\tilde{A}}_A)$ which, unless the states
are the same (up to a phase), is less than one. 
For two differing ground states, these infinite sums thus become geometric 
series $\sum_{n=0}^\infty (e^{+\ic p} E^A_{\tilde{A}})^n = (\mathbb{I} - 
e^{+\ic p} E^A_{\tilde{A}})^{-1}$.
If the states are the same, then $\rho(E^A_{\tilde{A}}) = 1$ and the inverses
must be replaced by pseudo-inverses. 
For example, a part of $B^\dagger \mathbf{H}_p B'$ where all three summed-over 
lattice sites are separated is
\begin{align*}
  \sum_{m=1}^{+\infty} \sum_{m'=1}^{+\infty}
    e^{+\ic pm} \braket{l|E^A_{B} (E^A_{\tilde{A}})^{m-1} E^{B'}_{\tilde{A}} (E^{\tilde{A}}_{\tilde{A}})^{m'-1} H^{\tilde{A}\tilde{A}}_{\tilde{A}\tilde{A}}|\tilde{r}} \\
  = e^{+\ic p} \braket{l|E^A_B (\mathbb{I} - e^{+\ic p} E^A_{\tilde{A}})^{-1} E^{B'}_{\tilde{A}} (\mathbb{I} - E^{\tilde{A}}_{\tilde{A}})^\mathrm{P} H^{\tilde{A}\tilde{A}}_{\tilde{A}\tilde{A}}|\tilde{r}},
\end{align*}
where $\tilde{r}$ is the right eigenvector of $E^{\tilde{A}}_{\tilde{A}}$,
$m$ is the number of sites between $B$ and $B'$, $m'$ is the number
of sites between $B'$ and the Hamiltonian term $h$ and we assume 
$\rho(E^A_{\tilde{A}}) < 1$. 
The Hamiltonian term is contained within $H^{\tilde{A}\tilde{A}}_{\tilde{A}\tilde{A}}$:
\begin{align*}
  H^{AB}_{CD} = \sum_{stuv} \braket{st|h - \braket{h}|uv} A^s B^t \otimes \overline{C^u D^v}.
\end{align*}
There is no infinite term corresponding to the pseudo-inverse in the above example 
because $\braket{\tilde l|H^{\tilde{A}\tilde{A}}_{\tilde{A}\tilde{A}}|\tilde{r}} = 0$.
Additional simplifications can be made by implementing the gauge-fixing condition
\begin{align*}
  \sum_s B^s \tilde{r} \tilde{A}^{s\dagger} = 0 \quad \iff \quad E^B_{\tilde{A}}\ket{\tilde{r}} = 0.
\end{align*}
A corresponding parametrization of $B^s$ is
\begin{align}
  B^s(x) = l^{-1/2} x \tilde{V}^s \tilde{r}^{-1/2},
  \label{eq:mps_tdvp_uni_B_param_excite}
\end{align}
where $x \in M_{D \times D(d - 1)}$ and the $D(d - 1) \times dD$  matrix 
$[\tilde{V}]_{(\alpha,s);\beta} = [\tilde{V}^s]_{\alpha\beta}$
is defined so that $\tilde{V}^\dagger$ contains an orthonormal basis 
($\tilde{V}\tilde{V}^\dagger = \mathbb{I}$) for the null-space of $\tilde R^\dagger$, with
\begin{align*}
  [\tilde{R}]_{(\alpha, s); \beta} = [\tilde{r}^{1/2} \tilde{A}^{s\dagger}]_{\alpha, \beta},
\end{align*}
resulting in $\tilde{V}\tilde{R} = 0$. For $\tilde{A} = A$, this parametrization 
is identical to \eqref{eq:mps_tdvp_uni_B_param}.
With it, the overlap term becomes 
$B(x)^\dagger \mathbf{N}_p B(y) = \tr[x^\dagger y] = \braket{x|y}$ 
so that the problem turns into a standard eigenvalue problem.
The effective Hamiltonian term becomes
\begin{align}
  &B^\dagger(x) \mathbf{H}_p B(y) = 
    \braket{l|H^{B(y)\tilde{A}}_{B(x)\tilde{A}}|\tilde{r}}
    + \braket{l|H^{AB(y)}_{AB(x)}|\tilde{r}} \nonumber \\
   &+ e^{+\ic p} \braket{l|H^{AB(y)}_{B(x)\tilde{A}}|\tilde{r}}
    + e^{-\ic p} \braket{l|H^{B(y)\tilde{A}}_{AB(x)}|\tilde{r}} \nonumber \\
   &+ \braket{l|E^{B(y)}_{B(x)} (\mathbb{I} - E^{\tilde{A}}_{\tilde{A}})^\mathrm{P} H^{\tilde{A}\tilde{A}}_{\tilde{A}\tilde{A}}|\tilde{r}}
    + \braket{l| H^{AA}_{AA} (\mathbb{I} - E^A_A)^\mathrm{P} E^{B(y)}_{B(x)}|\tilde{r}} \nonumber \\
   &+ e^{+\ic p} \braket{l|E^A_{B(x)}(\mathbb{I} - e^{+\ic p} E^A_{\tilde{A}})^{-1} E^{B(y)}_{\tilde{A}} (\mathbb{I} - E^{\tilde{A}}_{\tilde{A}})^\mathrm{P} H^{\tilde{A}\tilde{A}}_{\tilde{A}\tilde{A}}|\tilde{r}} \nonumber \\ 
   &+ e^{-\ic p} \braket{l|E^{B(y)}_{A} (\mathbb{I} - e^{-\ic p} E_A^{\tilde{A}})^{-1} E^{\tilde{A}}_{B(x)} (\mathbb{I} - E^{\tilde{A}}_{\tilde{A}})^\mathrm{P} H^{\tilde{A}\tilde{A}}_{\tilde{A}\tilde{A}}|\tilde{r}} \nonumber \\
   &+ e^{+\ic p} \braket{l|E^A_{B(x)}(\mathbb{I} - e^{+\ic p} E^A_{\tilde{A}})^{-1} H^{B(y)\tilde{A}}_{\tilde{A}\tilde{A}}|\tilde{r}} \nonumber \\
   &+ e^{-\ic p} \braket{l|E^{B(y)}_{A} (\mathbb{I} - e^{-\ic p} E_A^{\tilde{A}})^{-1} H^{\tilde{A}\tilde{A}}_{B(x)\tilde{A}}|\tilde{r}} \nonumber \\
   &+ e^{+2\ic p} \braket{l|E^A_{B(x)}(\mathbb{I} - e^{+\ic p} E^A_{\tilde{A}})^{-1} H^{AB(y)}_{\tilde{A}\tilde{A}}|\tilde{r}} \nonumber \\
   &+ e^{-2\ic p} \braket{l|E^{B(y)}_{A} (\mathbb{I} - e^{-\ic p} E_A^{\tilde{A}})^{-1} H^{\tilde{A}\tilde{A}}_{AB(x)}|\tilde{r}}, \label{eq:mps_umps_excite_eff_H}
\end{align}
where we again note that the inverses turn to pseudo-inverses if $A = \tilde{A}$.
It is possible to implement these operations with $\mathcal{O}(D^3)$ time 
complexity, avoiding direct calculation of inverses as in the TDVP algorithm
(see section \ref{sec_mps_umps_tdvp}).
A sparse eigenvalue solver can then be used to efficiently obtain eigenvalues.

A final ingredient is needed to define the momentum $p$ in the case 
$A \neq \tilde{A}$, because an overall phase on $A$
effectively shifts the momentum of the ansatz states
\begin{align*}
  \ket{\Phi_p(B;e^{\ic \phi}A,e^{\ic \varphi}\tilde{A})} \sim \ket{\Phi_{p + \phi - \varphi}(B;A,\tilde{A})},
\end{align*}
which can be seen in \eqref{eq:mps_umps_excite_eff_H}, where every factor $A$ ($\tilde A$)
is paired either with $A^\dagger$ ($\tilde A^\dagger$) (cancelling any extra phase factor) 
or with $e^{+\ic p}$ ($e^{-\ic p}$) (resulting in the momentum shift). 
We adhere to the convention of \cite{haegeman_variational_2012} and demand that 
the largest eigenvalue of $E_A^{\tilde{A}}$ is real and positive which,
in the case of equivalent states differing only by a phase $A = e^{\ic \phi}\tilde A$,
corresponds to $\phi = 0$.

\subsubsection{Mean-field case}
\label{sec_mps_MFT_excitations}

In the mean field case $D=1$, where there is no inter-site entanglement, the uMPS 
excitation ansatz simplifies further. The trial states are
\begin{align*}
  &\ket{\Phi_p(\vec{b};\vec{a},\tilde{\vec{a}})} = \\
     & \qquad \sum_{n \in \mathbb{Z}} e^{\ic pn}
                     \left[ \bigotimes_{-\infty}^{n-1} \ket{\psi(\vec{a})} \right]
                     \otimes \ket{\psi(\vec{b})} \otimes
                     \left[ \bigotimes_{n+1}^{+\infty} \ket{\psi(\tilde{\vec{a}})} \right],
\end{align*}
where the rank 3 tensors $A$ and $B$ of \eqref{eq:mps_excite_ansatz} have 
become vectors $\vec{a},\vec{b} \in \mathbb{C}^d$.
In this case, the $D^2 \times D^2$ operators $E$ and the corresponding vectors 
are just numbers so that the requirement
$\rho(E^A_A) = 1$ with the largest eigenvalue being $1$
implies $E^A_A = \vec{a}\cdot\vec{a} = 1$. The normalized ``eigenvectors'' 
$\bra{l}$ and $\ket{r}$ are thus also equal to 1 and projecting them out 
using $Q = 1 - \ket{r}\bra{l}$ 
leaves zero. All terms in \eqref{eq:mps_umps_excite_eff_H} containing the 
pseudo-inverse of $\mathbb{I} - E^A_A$ or 
$\mathbb{I} - E^{\tilde{A}}_{\tilde{A}}$ thus drop out. In the case
 $A = \tilde{A}$, this leaves only the first four terms.
Otherwise the inverse factors $(\mathbb{I} - e^{+\ic p} E^A_{\tilde{A}})^{-1}$
and $(\mathbb{I} - e^{-+\ic p}  E_A^{\tilde{A}})^{-1}$ are just positive numbers
and the last four terms are non-zero as well.

The gauge-fixing condition $\sum_s B^s \tilde{A}^{s\dagger} = 0$ corresponds
to the elimination of the norm degree of freedom, where the parametrization
\eqref{eq:mps_tdvp_uni_B_param_excite} forces $\vec{b}$ into the subspace 
orthogonal to $\tilde{\vec{a}}$. We can directly obtain the effective 
Hamiltonian as a $d \times d$ matrix
\begin{align}
  [\mathbf{H}_p]_{s,t} &= \braket{s \tilde{\psi}|h'|t \tilde{\psi}}
    + \braket{\psi s|h'|\psi t} \nonumber \\
    &+ e^{+\ic p} \braket{s \tilde{\psi}|h'|\psi t}
    + e^{-\ic p} \braket{\psi s|h'|t \tilde{\psi}} \nonumber \\
    &+ \left[ e^{+\ic p} \braket{s|\psi} (1 - e^{+\ic p}\braket{\tilde{\psi}|\psi})^{-1} \braket{\tilde{\psi}\tilde{\psi}|h'|t \tilde{\psi}} \right. \nonumber \\
    &+ e^{-\ic p} \braket{\psi|t} (1 - e^{-\ic p}\braket{\psi|\tilde{\psi}})^{-1} \braket{s \tilde{\psi}|h'|\tilde{\psi}\tilde{\psi}} \label{eq:mps_MFT_excite_effH}\\
    &+ e^{+2\ic p} \braket{s|\psi} (1 - e^{+\ic p}\braket{\tilde{\psi}|\psi})^{-1} \braket{\tilde{\psi}\tilde{\psi}|h'|\psi t} \nonumber \\
    &+ \left. e^{-2\ic p} \braket{\psi|t} (1 - e^{-\ic p}\braket{\psi|\tilde{\psi}})^{-1} \braket{\psi s|h'|\tilde{\psi}\tilde{\psi}} \right], \nonumber
\end{align}
where $\ket{\tilde{\psi}} \equiv \ket{\psi(\tilde{\vec{a}})}$ and 
$h' = h - \braket{h}$. In the topologically trivial case,
where $\vec{a} = \tilde{\vec{a}}$, the terms in square brackets drop out due to
$\vec{b} \cdot \tilde{\vec{a}} = \vec{b} \cdot \vec{a}= 0$.

\section{Studying quantum fields with matrix product states}
\label{chap:phi4}
In this section, we use the variational conjugate-gradient method for 
uniform matrix product states (uMPS) of section \ref{sec:mps_CG}
to determine the continuum critical parameter of $\phi^4$-theory, improving
on previous numerical results.
We also study the special case of uniform mean-field theory (MFT) states,
which correspond to uMPS with bond dimension one.
As well as the vacuum expectation value of the field, which plays the role of the 
order-parameter (see section \ref{sec:phi4_SSB}), we investigate the 
energy of the lowest-lying excitation as a phase-change indicator, which tends 
to zero at the critical point and, by universal correspondence to the Ising model, 
should scale linearly in its vicinity.
Furthermore, we extract the central charge of the conformal 
field theory (CFT) of the critical system \cite{pollmann_theory_2009},
which is also expected to be universal
\cite{latorre_ground_2003, calabrese_entanglement_2004, vidal_entanglement_2003}
and calculate the spectral density function of the near-critical lattice theory.

\subsection{Method}

As set out in section \ref{sec:qft_phi4_lattice}, $(1+1)$ dimensional 
$\phi^4$-theory can be put on a spatial lattice, in a way that formally 
recovers the continuum theory in the limit of zero lattice spacing 
$a \rightarrow 0$, using the nearest-neighbor Hamiltonian
\begin{align*}
  \tilde{H} = \sum_n \left[ \frac{\pi_n^2}{2}
      + \frac{(\phi_n - \phi_{n+1})^2}{2}
      + \frac{\tilde{\mu}_0^2}{2} \phi_n^2 + \frac{\tilde{\lambda}}{4!} \phi_n^4 \right],
\end{align*}
where $\tilde{\lambda} \equiv \lambda a^2$ and $\tilde{\mu}_0^2 \equiv \mu_0^2 a^2$ 
are dimensionless parameters. 
The theory exhibits spontaneous symmetry-breaking, as detailed in
section \ref{sec:phi4_SSB}, where a particular value of 
$\tilde{\lambda} / \tilde{\mu}_R^2(\tilde\lambda)$ characterizes the critical point for
a particular lattice-spacing $a$, hence the dependency on $\tilde\lambda(a)$. 
$\tilde{\mu}_R^2 = \tilde{\mu}_0^2 + \delta \tilde{\mu}^2_1(\tilde{\mu}_R^2)$ is the renormalized mass,
which is finite for $a > 0$ and is given by \eqref{eq:qft_phi4_lattice_mass_shift}. 
We use the uMPS conjugate gradient algorithm of section
\ref{sec:mps_CG} to obtain ground states up to some tolerance $\eta$
(see section \ref{sec_tdvp_imtime})
giving us access to approximate
ground-state expectation values, and the uMPS excitation ansatz of section 
\ref{sec:mps_uniform_excitations} to obtain excitation energies.
To study the system using uMPS, we first need an appropriate basis.

\subsubsection{Position basis with a cut-off}
\label{sec:phi4mps_basis}
To represent states using the uMPS formalism we choose the position basis described
in section \ref{sec:qft_phi4_lattice}:
\begin{align} \label{phi4mps_basis}
  &\ket{s_n} = \frac{(a_n^\dagger)^s}{\sqrt{s!}}\ket{0_n}
  \qquad [a_n, a_m^\dagger] = \delta_{nm} \\
  &\phi_n = \frac{1}{\sqrt{2}} \left( a_n^\dagger + a_n \right)
  \qquad \pi_n = \frac{\ic}{\sqrt{2}} \left( a_n^\dagger - a_n \right).  
\end{align} 
We provide the matrix-elements of relevant operators for this basis in appendix
\ref{apn_phi4_matrix_elements}.
Since the site subspace is infinite, we must introduce a cut-off so that states
can be stored using a finite number of parameters. We therefore limit ourselves 
to $\mathcal{H}_n = \mathbb{C}^d$ such that the highest available 
number-eigenstate is $\ket{d - 1}$, assuming that a good approximation to the
ground state does not require the higher modes to be present.
That this should be the case for the symmetric phase seems intuitive
considering the form of the classical effective potential (see Figure 
\ref{fig:phi4_Veff_cl}), but things are less clear for
the symmetry-broken case where the ground state is centered about one of two
separated wells away from the origin. The cut-off may thus affect the accuracy
of symmetry-broken states more significantly than symmetric ones. 
We also expect the higher modes to be more important 
for states near to the critical point, where fluctuations diverge.

\subsubsection{Field-shifted basis}

\begin{figure}
    \includegraphics{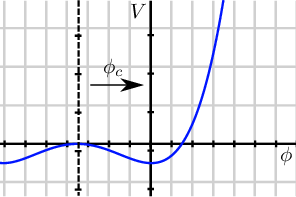}
    \caption{Illustration of the field shift needed to center fluctuations 
             about zero.}
    \label{fig:phi4mps_phi4_pot_shift}
\end{figure}
It should be possible to avoid higher excitations in the symmetry-broken 
phase $\braket{\phi} \neq 0$, thus mitigating the effects of the cut-off,
by changing the basis such that the operator $\phi'$ in the new basis
has an expectation value of approximately zero $\braket{\phi'} \approx 0$.
We effectively shift the origin in a plot of the effective potential by some
amount $\phi_c$ towards the minimum, such that fluctuations are centered about 
$\phi' = 0$. That higher excitations in the shifted number basis are then 
avoided seems intuitively reasonable given the classical effective potential, 
where each of the two wells (in the symmetry-broken case) looks locally
similar to a single-well potential. Figure \ref{fig:phi4mps_phi4_pot_shift}
illustrates this procedure.

The change of basis corresponds to the unitary 
\begin{align*}
  U(\phi_c) = e^{\ic \phi_c \pi},
\end{align*}
with $\pi$ being the conjugate momentum operator of \eqref{phi4mps_basis}.
It defines new creation and annihilation operators 
\begin{align*}
  a' = a - \frac{\phi_c\sqrt{2}}{2}
\end{align*}
such that $\phi'_n = \phi_n - \phi_c$ where $\phi_c \in \mathbb{R}$ characterizes 
the shift. In terms of operators in the shifted basis, the Hamiltonian
is
\begin{align*}
  \tilde{H} = \sum_n &\left[ \frac{{\pi'}_n^2}{2}
      + \frac{(\phi'_n - \phi'_{n+1})^2}{2} \right. \\
      &+ \frac{\tilde{\mu}_0^2}{2} ({\phi'}_n^2 + 2\phi_c{\phi'}_n + \phi_c^2) \\
      &\left. + \frac{\tilde{\lambda}}{4!} ({\phi'}_n^4 + 4 \phi_c {\phi'}_n^3 + 
                                    6\phi_c^2 {\phi'}_n^2 + 4 \phi_c^3 {\phi'}_n 
                                    + \phi_c^4) \right].
\end{align*}
Using this Hamiltonian with a value of $\phi_c \approx \braket{\phi}$ should
thus help to avoid higher excitations and allow us to efficiently
represent ground states with large values of $|\braket{\phi}|$.

\subsubsection{Effects of the Hilbert space cut-off}

\begin{figure}
  \includegraphics{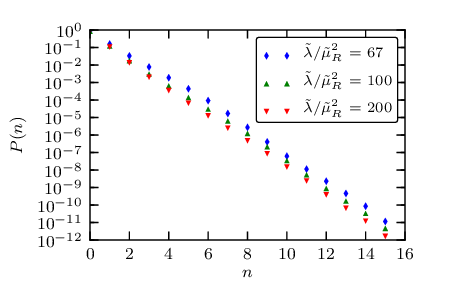}
    \vspace{-0.5cm}
  \caption{
    Histogram plots for the number operator in the shifted basis at varying
    distances $\tilde{\lambda} / \tilde{\mu}_R^2 = 67, 100, 200$ from the 
    critical point $\tilde{\lambda} / \tilde{\mu}_{R,c}^2 \approx 66$
    in the symmetry-broken phase 
    (with $\tilde{\lambda} = 0.1$ and $D = 128, 64, 64$ respectively). The higher
    modes carry more weight for states nearer the critical point.
  }
  \label{fig:shift_effect_phi_hist_near_crit}
\end{figure}

\begin{figure}
  \begin{center}
  \includegraphics{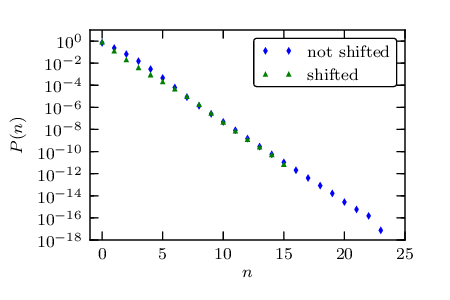}
    \\ \vspace{-0.4cm}
  \includegraphics{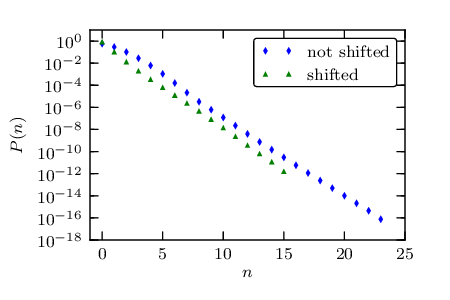}
  \end{center}
  \vspace{-0.8cm}
  \caption{
    Histogram plots for the number operator in the shifted and non-shifted
    bases near (top, $\tilde{\lambda} / \tilde{\mu}_R^2 = 80$) and far from 
    (bottom, $\tilde{\lambda} / \tilde{\mu}_R^2 = 200$) the critical point in the 
    symmetry-broken phase ($\tilde{\lambda} = 0.1$).
    The effect of shifting by approximately $\braket{\phi}$ is much stronger
    far into the symmetry-broken region (higher $\braket{\phi}$), where
    we see a weight-shift from the higher modes to the zero mode.
    The shifted states were obtained with $d = 16$, the non-shifted with 
    $d = 24$, hence the greater range of the non-shifted points.
    All four states have $D = 64$.
  }
  \label{fig:shift_effect_n_hist}
\end{figure}

\begin{figure}
    \includegraphics{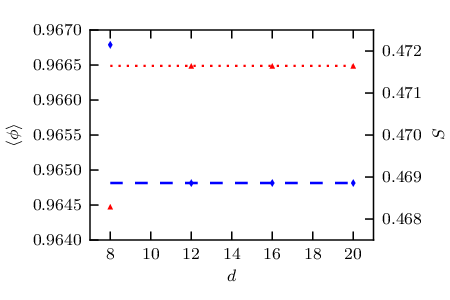} \vspace{-0.2cm} \\
    \includegraphics{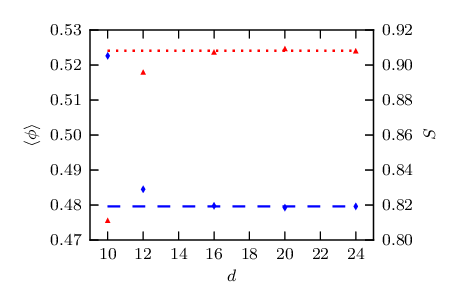}
    \vspace{-0.4cm}
  \caption{
    Scaling of $\braket{\phi}$ (blue diamonds) and the half-chain entropy $S$ (red triangles) with 
    the Hilbert space cut-off $d$ far into the symmetry-broken
    phase (top, $\tilde{\lambda} / \tilde{\mu}_R^2 = 200$, $D=64$) and near
    to the critical point (bottom, $\tilde{\lambda} / \tilde{\mu}_R^2 = 67$, $D=128$)
    using a shifted basis. In both cases $\tilde{\lambda} = 0.1$. It appears that 
    $d = 16$ is sufficient both near and far from the critical point. Additional 
    variation for $d \ge 16$ in the near-critical case is due to high
    sensitivity to the level of convergence (states were obtained with a 
    tolerance of $\eta < 3 \cdot 10^{-7}$). 
  }
  \label{fig:smalld_scaling}
\end{figure}

The effects of the local Hilbert space cut-off $d$ are, as expected, relatively
strong near to the critical point, becoming weaker further into the symmetry-broken
phase when using the shifted basis (see Figure \ref{fig:shift_effect_phi_hist_near_crit}). 
Without the basis shift, states with large
values of $\braket{\phi}$ exhibit a weight-shift towards higher modes, as illustrated in Figure 
\ref{fig:shift_effect_n_hist}.
In all cases, excitation of higher modes drops off exponentially, 
with $d = 16$ being sufficient to capture the most significant contributions, 
as demonstrated in Figure \ref{fig:smalld_scaling}.

\begin{figure}
  \includegraphics{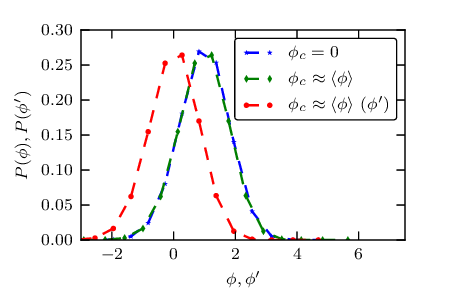}
  \vspace{-0.2cm}
  \caption{
    Visualization of the basis shift far into the symmetry-broken
    phase ($\tilde{\lambda} = 0.1$, $\tilde{\lambda} / \tilde{\mu}_R^2 = 200$). 
    $\phi$ histograms are plotted with (green diamonds) and without (blue stars) the basis
    shift $\phi_c \approx \braket{\phi}$. 
    Additionally, the histogram of the shifted-basis operator $\phi'$ is 
    plotted for the shifted state (red dots).
    States were obtained with $d = 16, D = 64$.
  }
  \label{fig:shift_effect_phi_hist}
\end{figure}

The shifted basis has an added benefit when sweeping 
$\tilde{\lambda} / \tilde{\mu}_R^2$ in the broken phase and using the previous
ground state approximation as a starting state for the next ground-state search. 
In this case, adjusting the shift $\phi_c$ towards the next predicted $\braket{\phi}$ 
(according to a preliminary fit of \eqref{eq:phi4mps_crit_phi}) improves the 
starting state by bringing $\braket{\phi}$ closer to the new ground-state value, 
leading to faster convergence. This is because
a shift of $\braket{\phi}$ always centers the state about the origin in 
the shifted basis (see Figure \ref{fig:shift_effect_phi_hist}).
Adjusting the shift by some $\Delta \phi$ also adjusts $\braket{\phi}$ by 
the same amount.

\subsubsection{Locating the critical point using the field expectation value}
\label{sec:phi4mps_locate_cp}

As noted in section \ref{sec:phi4_SSB}, since $\braket{\phi}$ is the 
order-parameter associated with the $\phi^4$-theory phase-change, it can be used 
to identify the critical point. A possible strategy for finding 
the critical parameters for $a,\tilde\lambda > 0$ might thus 
be to fix $\tilde{\lambda}$ and sweep $\tilde{\mu}_R^2$ until one sees a 
transition from $\braket{\phi} \neq 0$ to 
$\braket{\phi} = 0$ or vice versa. 
However, this is not practical because the amount of entanglement in the ground
state (for example, as quantified by the half-chain entropy \eqref{eq:umps_S_hc}) 
tends to infinity as
the critical point is approached, such that accurate representation using uMPS
requires the bond-dimension $D$ to approach infinity also. Since the 
computational complexity of the TDVP algorithm scales as $\mathcal{O}(D^3)$,
this bisection method cannot achieve high accuracy for reasons of practicality.

Instead, we approach the critical point from the symmetry-broken phase, noting
that physical quantities obey power laws in the vicinity of critical points 
(see section \ref{sec:phi4_SSB}). For $\braket{\phi}$ we can thus write
\begin{align}
  \braket{\phi} = A(\tilde{\lambda}) \left[ 
             \frac{\tilde{\lambda}}{\tilde{\mu}_R^2}
             - \frac{\tilde{\lambda}}{\tilde{\mu}_{R,c}^2(\tilde{\lambda})}
             \right]^{\beta(\tilde{\lambda})},
  \label{eq:phi4mps_crit_phi}
\end{align}
where $A(\tilde{\lambda})$ is a constant, $\beta(\tilde{\lambda})$ is the 
critical exponent and $\tilde{\mu}_{R,c}^2(\tilde{\lambda})$ is the 
critical value of $\tilde{\mu}_R^2$ for a given $\tilde{\lambda}$. 
Fitting this equation to $\braket{\phi}$ as a function of 
$\tilde{\lambda}/\tilde{\mu}_R^2$ (with fixed $\tilde \lambda$) as near as
possible to the phase-transition,
we obtain an estimate for the lattice critical parameter 
$\tilde \lambda / \tilde{\mu}_{R,c}^2(\tilde \lambda)$. We can then
use a series of fits with $\tilde{\lambda} \rightarrow 0$ to extrapolate
an estimate for the critical parameter $\lambda / \mu_R^2$ of the continuum theory. 

\begin{figure}[h]
  \includegraphics{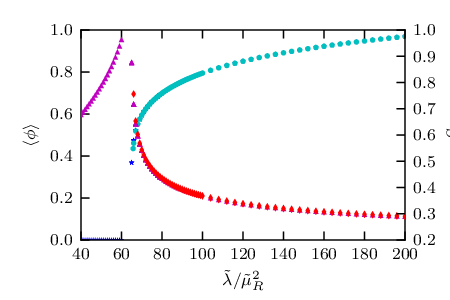}
  \caption{
    An example plot of the order parameter $\braket{\phi}$ (dots and stars) for 
    fixed $\tilde{\lambda} = 0.5$, sweeping $\tilde{\lambda} / \tilde{\mu}_R^2$. 
    The half-chain entropy $S$ is also shown (triangles and diamonds). 
    The cyan and red points (dots and diamonds) represent high bond-dimension limits with 
    $D \le 80$, whereas the blue and magenta points (stars and triangles) are for fixed $D = 32$.
    All ground state approximations are converged to a state tolerance 
    $\eta < 10^{-6}$.
  }
  \label{fig:sweep_phi_example}
\end{figure}

\begin{figure}[h] 
  \includegraphics{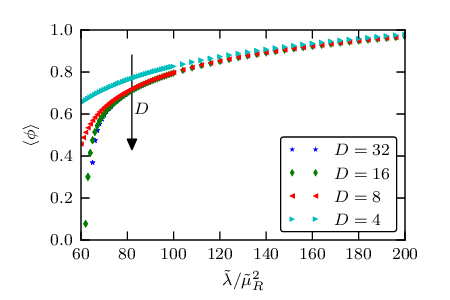}
  \caption{
    A plot of $\braket{\phi}$ for fixed  $\tilde{\lambda} = 0.5$, 
    sweeping $\tilde{\lambda} / \tilde{\mu}_R^2$ for several bond dimensions.
    At lower values of $D$, finite-entanglement effects shift the apparent 
    location of the critical point to lower
    values of $\tilde{\lambda} / \tilde{\mu}_R^2$.
  }
  \label{fig:D_scaling_sweep_phi}
\end{figure}

\begin{figure}[h]
  \vspace{-0.5cm}
  \begin{center}
    \includegraphics{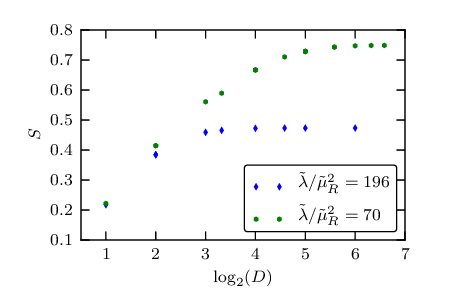}
    \includegraphics{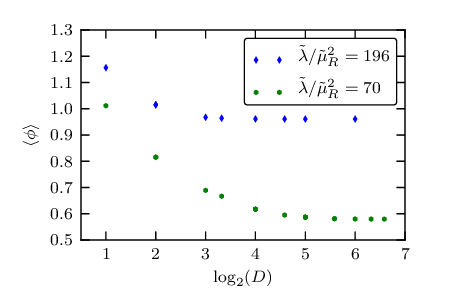}
  \end{center}
  \vspace{-0.5cm}
  \caption{
    Scaling of the half-chain entropy $S$ (top) and of $\braket{\phi}$ (bottom) 
    with the logarithm of the bond dimension $D$ for points near 
    (green hexagons, $\tilde{\lambda} / \tilde{\mu}_R^2 = 70$) 
    and far (blue diamonds, $\tilde{\lambda} / \tilde{\mu}_R^2 = 196$) from the critical 
    point ($\tilde{\lambda} = 0.1$).
    A higher bond dimension is necessary to accurately represent near-critical
    states compared to far-from-critical states.
  }
  \label{fig:D_scaling}
\end{figure}

\begin{figure}
    \begin{center}
    \includegraphics{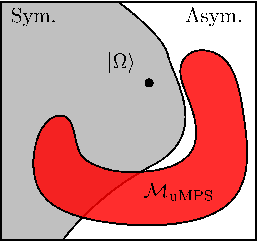}
    \vspace{-0.5cm}
    \end{center}
    \caption{
      Illustration of the relationship between the uMPS variational manifold
      for fixed bond-dimension $\mathcal{M}_\text{uMPS}$ and the ground state 
      $\ket{\Omega}$ for parameters close to the critical point in the symmetric 
      phase. Hilbert space is divided into symmetrical and asymmetrical 
      (in $\phi$) states.
    }
    \label{fig:phi4mps_sym_asym_M}
\end{figure}

Initial simulations show that, as expected, the half-chain entropy $S$ of the 
ground state approximation tends to infinity as the critical point is approached. 
This is visible in Figure \ref{fig:sweep_phi_example}, where we show results 
obtained from high bond dimension limits as well as using a fixed bond-dimension.
As further confirmed in Figures \ref{fig:D_scaling_sweep_phi} and 
\ref{fig:D_scaling}, a fixed $D$ is not sufficient
to capture ground states near the critical point.
Note that a phase transition does,
in fact, occur for fixed $D$, albeit not at the exact critical point, but at increasingly 
lower values of $\tilde{\lambda} / \tilde{\mu}_R^2$ for decreasing values of $D$ 
(and for decreasing $\tilde{\lambda}$).
This is consistent with the entropy shown in Figure \ref{fig:sweep_phi_example},
which is asymmetric about the critical point, falling off more slowly in the 
symmetric phase. Since a fixed $D$ represents an upper bound on the amount
of entanglement in the state (see section \ref{sec:mps_umps}), the uMPS 
variational manifold $\mathcal{M}_\text{uMPS}$ comes closer to the exact ground-state 
when its entropy $S$ is lower. 
Given the asymmetric entropy of $\phi^4$-theory, this implies that symmetry-broken 
ground-states are easier to approximate than symmetrical ones (for a given distance in parameter
space from the critical point). For a symmetric ground state with high entropy,
a low-lying excited state with much lower entropy may thus turn out to be the
best available ground state approximation in $\mathcal{M}_\text{uMPS}$.
Such an excitation should be available for such states, 
since a small change in the parameter $\tilde{\mu}_R^2$ results in an 
asymmetric ground state (on the other side of the critical point). The situation is 
illustrated in Figure \ref{fig:phi4mps_sym_asym_M}. The same $D$-dependent shift 
of the critical point is observed with the transverse Ising model \cite{tagliacozzo_scaling_2008}.

One might consider using this behavior together with finite-entanglement 
scaling techniques \cite{tagliacozzo_scaling_2008} to obtain information 
about the true critical point (for example, the critical exponent), but this 
requires precise knowledge of its location. Instead, we take data at several 
values of $D$ in order to obtain high-$D$ limits of $\braket{\phi}$, 
which we then fit using \eqref{eq:phi4mps_crit_phi} to obtain an estimate for
the location as well as the critical exponent.

Since we take a high-$D$ limit of the approximate ground-state value of
$\braket{\phi}$ for each parameter combination (requiring a higher $D$ for the 
higher-entropy states closer to the critical point) there is a practical limit on how near we 
can come. This is unfortunate, since the fit \eqref{eq:phi4mps_crit_phi}
is highly sensitive to near-critical points, where the gradient goes to infinity. 
Power-law scaling is also only exactly fulfilled 
infinitesimally close to the critical point, such that including data points
further away decreases accuracy. We thus only fit the points closest to the
critical point for which we have sufficient (in terms of bond-dimension) data.

\subsubsection{Locating the critical point using excitations}

Another approach to finding the critical parameters, given $\tilde{\lambda}$,
is to plot the energy of the lowest-lying excitation $\Delta \tilde E \equiv a \Delta E$ 
against $\tilde{\lambda}/\tilde{\mu}_R^2$, which 
should tend to zero as we approach the critical point 
$\tilde{\lambda}/\tilde{\mu}_{R,c}^2(\tilde\lambda)$ 
from either side. We describe an ansatz for obtaining the 
lowest-lying excitation energies, given a uMPS approximation to the 
ground state, in section \ref{sec:mps_uniform_excitations}. 

In obtaining excitation energies in the symmetry-broken phase, topologically non-trivial 
excitations must be taken into account. This precludes the use of the
$\phi$-shifted basis mentioned in section \ref{sec:phi4mps_basis} to represent the state, since 
approximations to both possible ground states are required for the calculations 
and these must use the same basis. For reasons of efficiency, it thus makes 
sense to focus on states near to the critical point where the shifted basis is 
not needed. This should not cause problems since this is where we expect
power-law scaling to be more exactly fulfilled.

\begin{figure}[h]
  \includegraphics{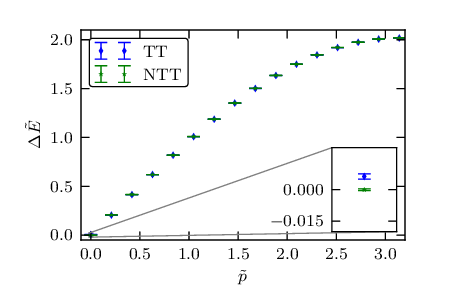}
  \caption{
    Extrapolated dispersion relation at the approximate lattice critical point
    $\tilde{\lambda}/\tilde{\mu}_R^2(\tilde{\lambda} = 1.0) \approx 64.4$
    showing the lowest-lying topologically trivial (TT) and topologically
    non-trivial (NTT) excitations. The zoomed area shows that the non-trivial
    excitation is the lowest-lying excitation at zero momentum. Momenta
    $0 \le p \le \pi/a$ are shown using $\tilde p \equiv ap$. 
    Energies $\Delta \tilde E$ are relative to the
    approximate ground-state energy. The bond-dimension is $D=32$ and
    points were linearly extrapolated from data at 
    $\tilde{\lambda}/\tilde{\mu}_R^2 = 70, 75, 80$.
  }
  \label{fig:disp_crit}
\end{figure}

\begin{figure}[h]
  \includegraphics{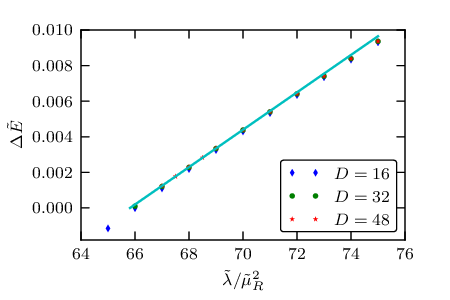}
  \caption{A parameter sweep of the energy of the lowest-lying excitation, 
  which is a soliton (with zero momentum) for $\tilde{\lambda} = 0.2$ at $D = 16, 32, 48$. 
  That the energy becomes negative 
  for low values of $\tilde{\lambda}/\tilde{\mu}_R^2$ indicates that these
  points lie in the symmetric phase (see main text). The line represents
  a fit to the data with $D=48$ using points 
  $\tilde{\lambda}/\tilde{\mu}_R^2 = 67 \dots 70$. The fitted value of
  the critical parameter is $\tilde{\lambda}/\tilde{\mu}_{R,c}^2 = 65.82(1)$.
  }
  \label{fig:gap_sweep}
\end{figure}

We first locate the lowest-lying excitation of the symmetry-broken phase 
and determine whether it is topologically trivial
or non-trivial, whilst confirming that it goes to zero for some value of $\tilde{\lambda}/\tilde{\mu}_R^2$.
To do this, we use the excitation ansatz to determine dispersion relations for 
the lowest-lying topologically trivial and non-trivial excitations for fixed 
$\tilde{\lambda}$ and several values of $\tilde{\lambda}/\tilde{\mu}_R^2$.
We then use linear extrapolation of the excitation energies at each 
momentum to obtain a dispersion relation at the first point where one of them goes to zero, 
which should correspond to the lattice critical point $\tilde{\lambda}/\tilde{\mu}_{R,c}^2(\tilde\lambda)$.
The result is shown in Figure \ref{fig:disp_crit}, where we see that the
lowest-lying excitations are the topologically non-trivial soliton (kink) excitations 
(at zero momentum). A plot of this excitation energy versus
$\tilde{\lambda}/\tilde{\mu}_R^2$ exhibits almost 
exactly linear scaling, consistent with the transverse Ising model, suggesting
the use of linear regression to obtain an estimate for the critical parameter. 
The plot is shown in Figure \ref{fig:gap_sweep}, which contains the excitation
energies obtained for several bond-dimensions. The small change in excitation energy
near the critical point when increasing the bond-dimension from $D=16$ to $D=48$, 
compared with the change in $\braket{\phi}$ shown in Figure \ref{fig:D_scaling_sweep_phi}
(for a larger lattice-spacing), suggests that finite-entanglement effects
are less severe for the excitation energy than for $\braket{\phi}$. Certainly,
the ground-state energy should reach a high-$D$ limit sooner than $\braket{\phi}$
simply because the approximate ground state is close to the energy minimum. Also, 
if the exact lowest-lying excitation is highly localized, we should need only 
a relatively low $D$ to approximate it well.
It seems excitations present a less computationally-intensive way of obtaining
a good estimate for the critical parameter, compared with $\braket{\phi}$.


We note that, for fixed $D$, the lowest-lying (soliton) excitation receives a 
negative energy 
(with respect to the approximate uniform ground-state) for sufficiently low
values of $\tilde{\lambda}/\tilde{\mu}_R^2$ (see Figure \ref{fig:gap_sweep}). 
The position of this crossover is very close to the critical point predicted by 
linear extrapolation. The existence of negative approximate excitation energies
indicates that the topologically non-trivial ansatz states include a better 
approximation to the exact ground state than $\mathcal{M}_\text{uMPS}$.
This is consistent with the $D$-dependent shift of the apparent phase-transition of
$\braket{\phi}$: If the exact ground state is symmetric, but the approximate ground
state is asymmetric, a topologically non-trivial excitation interpolating between 
the two degenerate asymmetric approximate ground states should be locally closer 
to the exact ground state at the disturbance. 

In fact, we can construct a better uMPS ground-state approximation $\ket{\Psi(A')}$ 
using negative-energy kink ``excitations'' by defining new $2D \times 2D$ 
parameter matrices
\begin{align*}
  A'^s = \begin{pmatrix}
         A^s & \epsilon B^s \\
         \epsilon \tilde B^s & \tilde A^s
       \end{pmatrix},
\end{align*}
where $A$ and $\tilde A$ are the parameters for the two original ground states,
 $B$ and $\tilde B$ are the tangent-vector parameters for a kink and an anti-kink
 and $\epsilon \in \mathbb{R}$.
The resulting state contains the original ground states as well as kink states
at order $\epsilon$ plus multi-kink states at $\mathcal{O}(\epsilon^2)$. This leads
to energy contributions $\epsilon^2 E_\text{kink}$ (there are 
no kink contributions at order $\epsilon$ because the kinks are orthogonal
to the original ground states). Since $E_\text{kink} < E_{\ket{\Psi(A)}}$ the
state $\ket{\Psi(A')}$ can, depending on $\epsilon$, 
have a lower energy than $\ket{\Psi(A)}$. $\epsilon$ can be interpreted
as a kink-density with an optimal value depending on the higher-order energy
contributions.

\subsubsection{Mean-field theory}
\label{sec:phi4mps_MFT}
When $D = 1$, the uMPS variational class is the same as that of the uniform 
product states (or mean field theory (MFT) states)
\begin{align*}
  \ket{\Psi(\vec{a})} = \dots \otimes \ket{\psi(\vec{a})} \otimes \ket{\psi(\vec{a})} \otimes \dots
\end{align*}
where $\ket{\psi(\vec{a})} = \sum_{s=0}^{d-1} a^s \ket{s}$ and 
$\vec{a} \in \mathbb{C}^d$. 
In this case, the per-site energy expectation value takes on an effective 
one-particle form
\begin{align*}
  \braket{h} &= \braket{\psi_n \psi_{n+1}|h_{n, n+1}|\psi_n \psi_{n+1}} \\
      &= \bra{\psi(\vec{a})} \left[ \frac{\pi^2}{2}
      + \frac{\tilde{\mu}_0^2}{2} \phi^2 + \frac{\tilde{\lambda}}{4!} \phi^4 \right]
      \ket{\psi(\vec{a})} + \sigma_\phi^2(\vec{a}),
\end{align*}
where $\sigma_\phi^2$ is the $\phi$-variance $\sigma_\phi^2(\vec{a}) = 
\braket{\psi(\vec{a})|\phi^2|\psi(\vec{a})} - 
\braket{\psi(\vec{a})|\phi|\psi(\vec{a})}^2$. 
An approximation to the ground state
can then be found by applying the time-independent variational principle and 
minimizing $\braket{h}$ with respect to the $d$ parameters $\vec{a}$,
which can be taken to be real since all matrix elements in the above expression
are real. The gradient of $\braket{h}$ is also readily obtainable
\begin{align*}
  \frac{\partial}{\partial \overline{\vec{a}}^s} \braket{h} 
      = 2 \bra{s} 
         &\left[ 
           \frac{\pi^2}{2} + \frac{\tilde{\mu}_0^2}{2} \phi^2 + \frac{\tilde{\lambda}}{4!} \phi^4
           + \phi^2 \right. \\
           &\quad\left. - 2\braket{\psi(\vec{a})|\phi|\psi(\vec{a})}\phi
           \vphantom{\frac{\tilde{\lambda}}{4!}}\right] \ket{\psi(\vec{a})},
\end{align*}
making many commonly-used minimizing algorithms applicable, such as the quasi-Newton 
method of Broyden, Fletcher, Goldfarb, and Shanno (BFGS) \cite{press_numerical_2007}. This
method is much simpler and more efficient than applying the
imaginary-time TDVP algorithm for uMPS with $D=1$.

Normalization presents a minor complication. The above equations assume 
$\braket{\psi(\vec{a})|\psi(\vec{a})} = 1$, which imposes a 
constraint $\vec{a}^2 = 1$ on the variational parameters.
Rather than using a constrained optimizer, we eliminate the norm degree of 
freedom by switching to $d$-dimensional spherical coordinates such that the norm
corresponds to a single parameter and can easily be fixed and ignored.

To estimate the location of the lattice critical point $\tilde\lambda/\tilde\mu_{R,c}^2$
using MFT, we again obtain ground 
states for a sweep of $\tilde\lambda/\tilde\mu_R^2$ for some fixed $\tilde\lambda$.
As in the more general case of uMPS with fixed $D$, we are putting a restriction
on entanglement by using MFT ($D=1$) and thus expect an apparent phase-transition 
to occur at some value of $\tilde\lambda/\tilde\mu_R^2 < \tilde\lambda/\tilde\mu_{R,c}^2$
for a given $\tilde\lambda$. We use the location of the apparent transition as an
estimate for $\tilde\lambda/\tilde\mu_{R,c}^2$, obtaining it from both $\braket{\phi}$
and excitation energies calculated using the MFT excitation ansatz of section 
\ref{sec_mps_MFT_excitations}. 
Since we do not expect power-law scaling of
physical quantities to be reproduced by MFT, we do not attempt to
fit data using power laws. Instead, we use bisection to pin down the
apparent phase-transition in $\phi$, which is possible due to the relative
ease of finding MFT ground states, and interpolate the lowest-lying excitation 
energies (in the apparently symmetry-broken phase) to obtain the point at which 
they become negative (these excitations are topologically non-trivial, as for $D > 1$
| see the above explanation).

Although we can obtain estimates for the lattice critical point using these
methods, we do not expect to obtain useful information about the continuum critical 
theory due to the lack of entanglement. 
However, since it is also possible to interpret the lattice critical 
point as a continuum limit of a non-critical theory (see section \ref{sec:phi4_SSB}), 
an ability to estimate its location using mean-field theory indicates 
that useful predictions about non-critical continuum theories can be made. 

\subsubsection{Central charge}
We determine the central charge associated with the conformal field theory
of the critical system using finite-entanglement scaling techniques.
It is known that, for infinite one-dimensional systems with a second-order 
phase-transition, the half-chain entropy of the ground state in the 
vicinity of a critical point with conformal invariance is
\begin{align}
  S = \frac{c}{6} \log (\xi / a),
  \label{eq:phi4mps_cft_S}
\end{align}
where $\xi$ is the correlation length and $c$ is the ``central charge''
\cite{calabrese_entanglement_2004}. Approaching the critical point, 
$\xi \rightarrow \infty$ and the entropy diverges.
The central charge specifies a conformal field theory (CFT), which 
describes behavior at the critical point in the continuum limit.

We know from \eqref{eq:umps_S_hc} that the maximum half-chain entropy of a 
uMPS state, contained, assuming right canonical form, in the diagonal entries
of the $D \times D$ matrix $l$, is directly related to the bond-dimension $D$,
which is also the maximum Schmidt-rank of the corresponding Schmidt 
decomposition. Thus, for lower values of $D$, finite-entanglement effects
occur and the value of $S$ scales with $D$.
It turns out there is a simple relationship between $S$, $D$, and $c$ describing
this scaling \cite{pollmann_theory_2009}
\begin{align}
  S = \frac{1}{\sqrt{12/c} + 1} \log D.
  \label{eq:phi4mps_cft_S_finite_D}
\end{align}
We can thus obtain an estimate for $c$ from values of $S$ taken from a 
number of ground state approximations with varying $D$. For this to work,
we must be close enough to the critical point so that \eqref{eq:phi4mps_cft_S} 
is valid and use small enough $D$ so that $S$ is limited by finite-entanglement
effects. We can then use linear regression to fit 
\eqref{eq:phi4mps_cft_S_finite_D} and obtain $c$.

\subsection{Results and analysis}

\subsubsection{Estimates of the continuum critical parameter}

\begin{figure}[h]
  \includegraphics{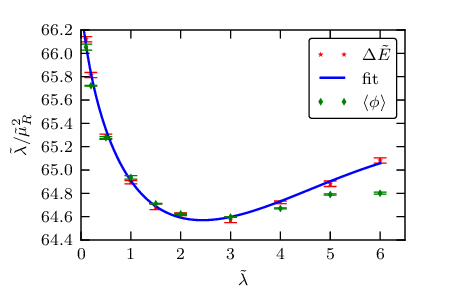}
  \vspace{-0.5cm}
  \includegraphics{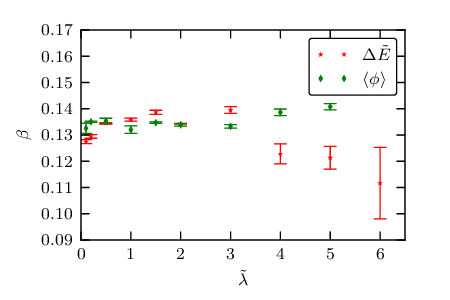}
      \caption{
        Approximate values for the lattice critical parameter 
        $\tilde{\lambda} / \tilde{\mu}_{R,c}^2(\tilde\lambda)$ (top) and the
        $\braket{\phi}$ critical exponent $\beta(\tilde\lambda)$ (bottom) obtained from linear fits 
        to the lowest-lying excitation energies $\Delta \tilde E$ and from power-law fits to the order-parameter $\braket{\phi}$ 
        for values of $\tilde{\lambda}$ approaching the continuum limit 
        $\tilde{\lambda} \rightarrow 0$. The line
        corresponds to the fourth fit of Table \ref{tab:phi4mps_crit_cont_lim_fits}.
      }
      \label{fig:crit_cont_lim}
\end{figure}

Figure \ref{fig:crit_cont_lim} shows estimates for the 
critical parameter $\tilde{\lambda} / \tilde{\mu}_{R,c}^2(\tilde\lambda)$ 
taken from sweep plots of $\braket{\phi}$ and of the lowest-lying excitation
energy $\Delta \tilde E$, 
approaching the continuum limit $\tilde{\lambda} \rightarrow 0$. 

\begingroup
\squeezetable
\begin{table}[h]
    \centering
    \begin{tabular}{|l|l|l|l|l|}
    \hline
    \hline
    & \multicolumn{2}{|c|}{$\braket{\phi}$} & \multicolumn{2}{|c|}{$\Delta \tilde E$} \\
    \hline
    Fit function & $f_c$ & $\chi^2/$dof & $f_c$ & $\chi^2/$dof \\
    \hline
    $f_c + c_1 \tilde{\lambda}$  & 65.10(18) & $9\times 10^3$ & 65.22(24) & 676 \\
    $f_c + c_1 \tilde{\lambda} + c_2 \tilde{\lambda}^2$  & 65.61(16) & $3 \times 10^3$ & 65.79(17) & 186 \\
    $f_c + c_1 \tilde{\lambda} + c_2 \tilde{\lambda} \ln \tilde{\lambda}$  & 66.01(11) & 771 & 66.19(11) & 44.8 \\
    $f_c + c_1 \tilde{\lambda} + c_2 \tilde{\lambda} \ln \tilde{\lambda} + c_3\tilde{\lambda}^2$ & 66.30(2) & 19.7 & 66.46(5) & 4.67 \\
    $f_c + c_1 \tilde{\lambda} + c_2 \tilde{\lambda} \ln \tilde{\lambda} + c_3\tilde{\lambda}^2 \ln \tilde{\lambda}$  & 66.26(3) & 24.8 & 66.42(5) & 6.22 \\
    \hline
    \hline
    \end{tabular}
    \caption{
      Fits, for lattice-spacings approaching zero, of the lattice critical parameter 
      $\tilde{\lambda} / \tilde{\mu}_{R,c}^2(\tilde\lambda)$ obtained from
      power-law fits to uMPS ground-state $\braket{\phi}$-values and from linear 
      extrapolation of lowest-level excitation energies $\Delta \tilde E$ (all in the symmetry-broken phase). 
      $f_c \equiv \lambda / \mu_{R,c}^2$ is the extrapolated continuum critical 
      parameter.
      We limit the $\braket{\phi}$ data fitted to obtain each 
      $\tilde{\lambda} / \tilde{\mu}_{R,c}^2$      
      to a few points close to the critical point with $\braket{\phi} \le 0.59$.
      The fitted data is plotted in Figure \ref{fig:crit_cont_lim}.
    }
    \label{tab:phi4mps_crit_cont_lim_fits}
\end{table}
\endgroup
The two sets of values show good
agreement, with the largest discrepancy occurring for $\tilde{\lambda} = 6$,
where we found high-$D$ limits of $\braket{\phi}$ particularly close
to the critical point without resorting to very high bond-dimensions.
Excluding the points of lowest $\braket{\phi}$ from the fit pushes the fitted 
value of $\tilde{\lambda} / \tilde{\mu}_{R,c}^2$ upwards, closer to the $\Delta \tilde E$ 
value, leading us to speculate that the excluded $\braket{\phi}$ values were not 
accurate enough, possibly due to insufficient convergence of the
uMPS ground state. We are inclined to trust the results obtained from the 
$\Delta \tilde E$ data over those from fits to $\braket{\phi}$, in particular due to the
relative robustness of the linear regression fit to errors made near the critical point.

As expected (see section \ref{sec:phi4_SSB}), non-linear behavior
of $\tilde{\lambda} / \tilde{\mu}_{R,c}^2(\tilde\lambda)$ is present. Given
that the exact behavior is unknown, but is predicted to be logarithmic, 
we follow \cite{schaich_improved_2009} and fit a series of 
functions, evaluating the $\chi^2$ statistic to judge which can be reasonably 
used to predict a continuum value $\lambda / \mu_{R,c}^2$. The results of the 
fits are listed in Table \ref{tab:phi4mps_crit_cont_lim_fits}, where we define
our final estimates for $\lambda / \mu_{R,c}^2$ to be
the fitted values with reduced $\chi^2$ statistic $\chi^2/$dof closest to one.

We find that the critical exponent $\beta(\tilde\lambda)$ obtained only from fits to
$\braket{\phi}$ agrees poorly with the predicted transverse Ising value 
of $0.125$, the fitted values near the continuum limit being significantly
higher, as shown in Figure \ref{fig:crit_cont_lim}.
This we attribute to insufficient data near to the lattice critical points, 
noting that the effect of excluding the points of lowest $\braket{\phi}$ 
is to increase the fitted value of $\beta(\tilde\lambda)$ further.
Using the $\tilde{\lambda} / \tilde{\mu}_{R,c}^2(\tilde\lambda)$ values taken 
from the $\Delta \tilde E$ data together with the $\braket{\phi}$ data, we obtain a 
second estimate of $\beta(\tilde\lambda)$ that, in the continuum limit 
$\tilde\lambda \rightarrow 0$, shows a much clearer trend towards the Ising value, 
in support of the greater reliability of the $\Delta \tilde E$-based estimates of the 
critical parameter.

\begin{figure}
  \includegraphics{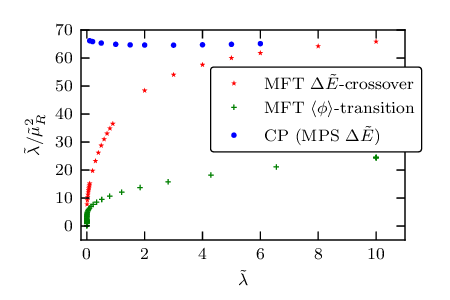}
  \caption{
    Position in $\tilde{\lambda}/\tilde{\mu}_{R}^2$ of the apparent
    phase-transition obtained from the mean-field results for $\braket{\phi}$
    (green crosses) and the lowest-lying excitation energy $\Delta \tilde E$ (red stars) 
    compared to the critical parameters
    $\tilde\lambda/\tilde\mu_{R,c}^2(\tilde\lambda)$
    obtained from the uMPS $\Delta \tilde E$ data (blue dots).
  }
  \label{fig:MFT_cont}
\end{figure}

\subsubsection{Mean-field results}

Due to the lack of entanglement in the mean-field class 
(see section \ref{sec:phi4mps_locate_cp}),
we expect a large shift of the apparent phase-transition to lower values of 
$\tilde{\lambda}/\tilde{\mu}_{R}^2$. 
We observe that the shift takes the apparent phase-transition 
(in $\braket{\phi}$ and in $\Delta \tilde E$) 
towards $\tilde{\lambda}/\tilde{\mu}_{R}^2 = 0$
in the continuum limit $\tilde{\lambda} \rightarrow 0$ (see Figure 
\ref{fig:MFT_cont}). For higher
values of $\tilde\lambda$, the apparent phase-transition starts to approach
the approximate critical parameters obtained above using uMPS with $D > 1$.
The mean-field excitations data gives better predictions for the critical
parameters than $\braket{\phi}$, in agreement with the relatively low sensitivity 
of the excitation energies to the bond-dimension observed with uMPS.

\subsubsection{Central charge}

\begin{figure}[h]
  \includegraphics{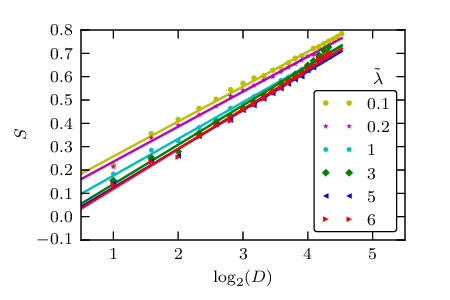}
  \includegraphics{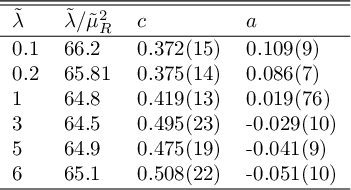}
  \caption{
    Entropy-scaling with $D$ for the near-critical lattice theory, using
    estimates for $\tilde\lambda/\tilde\mu_{R,c}^2(\tilde\lambda)$
    obtained from excitations data. The legend shows the value of $\tilde{\lambda}$.
    The table shows the parameters used and the values for the critical
    charge $c$ and the $S$-intercept $a$ derived from the linear fits.
  }
  \label{fig:phi4mps_cc}
\end{figure}

We extract the central charge by finding approximate uMPS ground states for
various low bond-dimensions at (or very near) the lattice critical point as 
determined from the uMPS excitations data. We observe approximately linear scaling of the 
entropy, as predicted by \eqref{eq:phi4mps_cft_S_finite_D}, with the central 
charge extracted from the gradient of a linear fit agreeing well,
for larger values of $\tilde{\lambda}$, with the prediction 
of $c=0.5$ from the transverse Ising model. 
For lower values of $\tilde{\lambda}$, we find that the gradient often agrees 
poorly with $c=0.5$, despite the scaling remaining linear. At this point,
we do not have a good explanation for this discrepancy, so we leave it as a
subject for future investigations. Our results are summarized in Figure
\ref{fig:phi4mps_cc}.

\subsubsection{Spectral density}

\begin{figure}
  \includegraphics{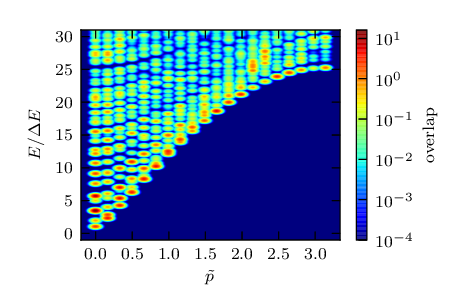}
  \caption{
    Spectral density function in the symmetry-broken phase 
    ($\tilde\lambda = 1$, $\tilde\lambda / \tilde\mu_R^2 = 77$) obtained at
    $D=16$. Dirac delta functions are replaced by Gaussians to aid visualization.
  }
  \label{fig:specdens}
\end{figure}

\begin{figure}
  \includegraphics{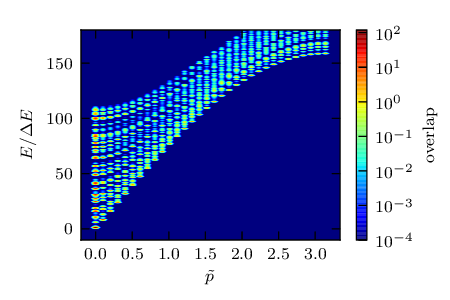}
  \caption{
    Spectral density function in the symmetric phase 
    ($\tilde\lambda = 1$, $\tilde\lambda / \tilde\mu_R^2 = 60$) obtained at
    $D=48$. Dirac delta functions are replaced by Gaussians to aid visualization.
  }
  \label{fig:specdens_TT}
\end{figure}

To further demonstrate the convenience of having approximate ground states in uMPS form
and the usefulness of the uMPS excitation ansatz,
we obtain the spectral density function \eqref{eq:qft_spec_dens} 
from the overlap of approximate
excited states with the approximate uMPS ground state. Results for states in the
symmetry-broken and in the symmetric phase are shown in Figure \ref{fig:specdens}
and Figure \ref{fig:specdens_TT} respectively. 
By interpreting the lattice critical point as a continuum limit of a non-critical
theory (see section \ref{sec:phi4_SSB}), a series of such plots could 
be used to extrapolate a continuum spectral density for that theory.
The nature of the uMPS excitation
ansatz means that not all excitations can be captured. This explains the lack
of a continuum of excitations, expected soon after the single-particle state. 
It would be interesting to compare the results with those of other approaches
such as \cite{osborne_renormalisation-group_2006}.

\subsection{Discussion}

The consistency of the lattice critical parameters obtained from approximate uMPS 
ground-state field expectation values $\braket{\phi}$ and those derived from 
the lowest-level excitation energies $\Delta \tilde E$ calculated using the uMPS excitation ansatz, 
as well as agreement of the critical exponents with their transverse Ising counterparts, 
demonstrates the validity of both methods for studying critical 
$(1+1)$-dimensional $\phi^4$-theory.
The finite-entanglement scaling method we use to obtain estimates for the
central charge $c$ of critical $\phi^4$-theory shows promise: Approximate linear
scaling of the entropy with $\log_2(D)$ is observed, as predicted, and the
fitted values for $c$ agree with the transverse Ising value $c=0.5$ for higher values
of $\tilde\lambda$. However, further work is needed to explain the discrepancies 
observed for lower $\tilde\lambda$.

We find uMPS to be an excellent class of ansatz states for studying the critical 
phenomena of $(1+1)$-dimensional $\phi^4$-theory due to the amount 
of entanglement in near-critical ground states being the main barrier to their efficient 
representation. With MPS, the amount of entanglement that can be represented is
controllable via the bond-dimension $D$, which can easily be varied to obtain
the limiting behavior of quantities such as $\braket{\phi}$. In this way, 
we can avoid errors originating from finite-entanglement effects. Additionally, 
working directly in the
thermodynamic limit of an infinite lattice completely avoids
additional finite-size effects and the need for further scaling investigations. 
Obtaining ground states using our variational conjugate gradient method 
(or the TDVP with imaginary time evolution) is also very convenient, 
since we need only enough storage capacity
to capture the approximate state at one point in time, unlike when simulating
the Euclidean theory on a space-time lattice. Also, using the TDVP, the computational
complexity scales linearly in $\tau$.

The class of uniform mean-field states (uMPS with $D=1$), despite featuring no 
entanglement, appears suited to estimating properties of non-critical continuum
theories in some cases, even if not of critical theories. Owing to the relative ease
with which MFT ground-state approximations and excitation energies can be obtained,
and the low computational cost of extending them to higher space-time dimensions
(due to the lack of entanglement),
these methods represent another useful tool for investigating quantum field theories.

\subsubsection{Comparison to other methods}

We summarize existing literature estimates for the continuum critical
parameter $\lambda/\mu^2_{R,c}$ in Table \ref{tab:phi4mps_crit_comparison},
where we include our results from Table \ref{tab:phi4mps_crit_cont_lim_fits}
with $\chi^2/$dof values closest to one. Our estimates agree
poorly with the DMRG result of \cite{sugihara_density_2004}, but relatively
well with the Monte Carlo results of \cite{schaich_improved_2009}.
With regard to the DMRG results, where the technique used is similar to ours,
we can attribute the large difference to finite-entanglement effects, since
these serve to shift the apparent critical point to lower values of
$\tilde{\lambda}/\tilde{\mu}^2_R$, with a larger shift for smaller
lattice-spacings. The DMRG parameters used in \cite{sugihara_density_2004}
correspond to $D=d=10$ \cite{rommer_class_1997}, resulting in a relatively
large shift (see Figure \ref{fig:D_scaling_sweep_phi}). The DMRG study
also uses only two lattice critical points to extrapolate a continuum value and,
as such, misses the non-linear behavior of $\tilde\lambda/\tilde\mu_{R,c}^2(\tilde\lambda)$.
The Monte Carlo methods used in \cite{schaich_improved_2009} are very
different to ours. They work with the Euclidean theory on a
finite two-dimensional lattice whereas we work on an infinite spatial
lattice in continuous time
(numerical integration of the TDVP flow equations 
could be seen as analogous to discretizing imaginary time on a lattice, in which
case our temporal ``lattice'' is of the length necessary to produce
sufficient convergence of the approximate ground-state). 
Rather than taking finite-size scaling
limits, we take finite-entanglement scaling limits to obtain our
ground-state approximations. Noting these differences, the fact that our results 
agree to within $2\%$ gives us confidence in the methods used.

\begin{table}[h]
    \centering
    \begin{tabular}{llll}
    \hline
    \hline
    Method & $f_c$ & Reference \\
    \hline
        uMPS, TDVP, $\Delta \tilde E$       &66.46(5) & This work \\
        uMPS, TDVP, $\braket{\phi}$  &66.30(2) & This work \\
        Monte Carlo	                 &$64.8^{+0.6}_{-0.3}$ & \cite{schaich_improved_2009}\\
        Gaussian effective potential &61.632 & \cite{hauser_connected_1995}\\
        Gaussian effective potential &61.266 & \cite{chang_existence_1976}\\
        GEP and oscillator rep.      &61.26  & \cite{ji_canonical_2002}\\
        Spherical field theory       &60.3   & \cite{lee_introduction_1998}\\
        Diffusion Monte Carlo        &$60 \pm 4.8 \pm 2.4$ & \cite{marrero_non-perturbative_1999}\\
        DMRG                         &59.89(1) & \cite{sugihara_density_2004}\\
        Continuum light-front        &59.46  & \cite{bender_spontaneous_1993}\\
        Connected Green function     &58.70  & \cite{hauser_connected_1995}\\
        Coupled cluster expansion    &$22.8 < f_c < 51.6$ & \cite{funke_approaching_1987}\\
        Discretized light-front      &43.95, 46.26  & \cite{sugihara_variational_1998}\\
        Discretized light-front      &43.70, 33.00  & \cite{harindranath_solving_1987, harindranath_stability_1988}\\
        Random phase approximation   &43.2   & \cite{hansen_random_2002}\\
        Non-Gaussian variational     &41.28	 & \cite{polley_second-order_1989}\\
    \hline
    \hline
    \end{tabular}
    \caption{
      Summary of results for the continuum critical parameter 
      $f_c \equiv \lambda/\mu_{R,c}^2$
      from the literature, including our results derived from lowest-lying 
      excitation energies $\Delta \tilde E$ and from $\braket{\phi}$, where we use the 
      results corresponding to the 
      $\chi^2/$dof values closest to one (see Table \ref{tab:phi4mps_crit_cont_lim_fits}).
    }
    \vspace{-1cm}
    \label{tab:phi4mps_crit_comparison}
\end{table}

\section*{Conclusion}

The class of uniform matrix product states (uMPS) appears well-suited to the
study of critical quantum fields in $(1+1)$ dimensions via lattice regularization.
Using variational methods like our naive variational conjugate gradient method
or the imaginary-time time-dependent variational principle (TDVP)
(where the former provides significantly improved convergence speed for the
system studied), 
good approximations to ground states can be obtained efficiently, even near
to the critical point. Here, the correspondence between the bond dimension $D$
and the maximum entanglement of a state allows the use of finite-entanglement
scaling to judge the accuracy of physical quantities calculated. 
Compared to Monte Carlo simulations, uMPS allows us to work directly in the 
thermodynamic limit and has storage requirements independent of the imaginary 
time dimension. Further, low-lying excitation energies are straightforward to
calculate, enabling the study of dispersion relations and the spectral density.
Even mean field theory shows potential for delivering useful predictions about 
non-critical continuum theories.

\emph{Acknowledgments} | Helpful discussions with Florian Richter and Cédric
Bény are gratefully acknowledged.
This work was supported by the ERC grant QFTCMPS and
by the cluster of excellence EXC 201 Quantum Engineering and Space-Time
Research.

\appendix
\section{Minimization using conjugate-gradient methods}
\label{sec:min_CG}

\begin{wrapfigure}{r}{3.2cm}
  \includegraphics[width=3cm]{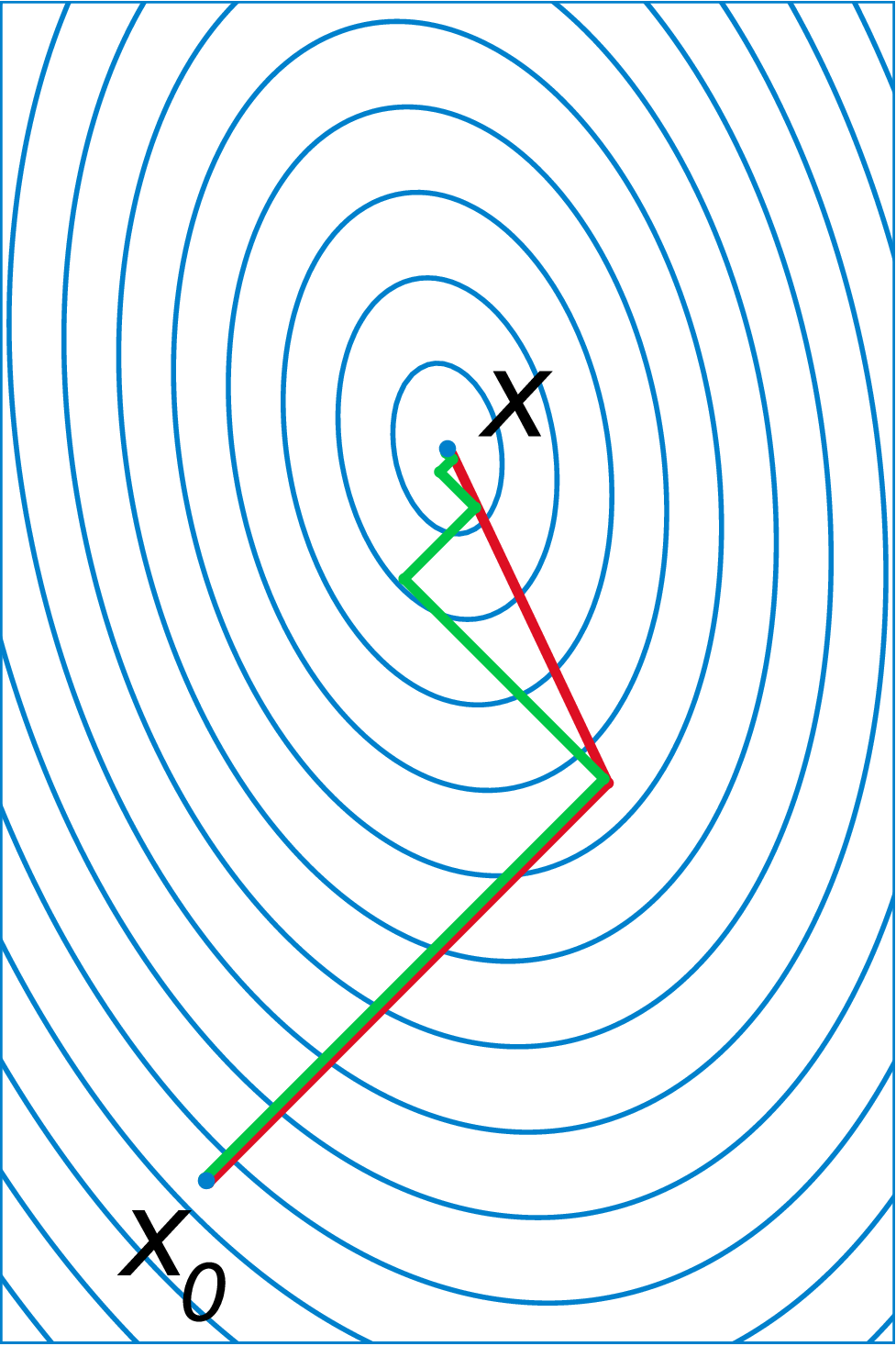}
\caption{
  Illustration \cite{cg_illustration} 
  of quadratic function (blue) minimization using
  the gradient descent (green) and conjugate gradient (red) methods.
        }
\end{wrapfigure}

When minimizing functions that are approximately quadratic 
in their parameters, which is always true in the vicinity of a minimum 
(assuming sufficient differentiability), making steps 
along the gradient direction (gradient-descent) is often a sub-optimal way of 
reaching the minimum.

We can illustrate this using a quadratic function of two variables $f(\vec{x})$
as shown in the figure to the right.
Beginning at some point $\vec{x}_0$ near the minimum and using a line-search to find
the minimum of $f$ in that direction to determine each step size (green line), 
an unfortunate starting position can result in a long zig-zag path and a large 
number of steps.

To avoid this, we can use the conjugate-gradient method \cite{press_numerical_2007}, 
which works by only stepping in directions that are conjugate to those already used.
Writing a general quadratic function of many variables as 
\begin{align*}
  f(\vec{x}) = \frac{1}{2} || \tilde{A}\vec{x} - \tilde{\vec{b}} ||^2 + \text{const.}
\end{align*}
with $\vec{x} \in \mathbb{C}^d$, $\tilde{A} \in M_{d \times d}$,
we can define $A = \tilde{A}^\dagger \tilde{A}$ and 
$\vec{b} = A^\dagger \tilde{\vec{b}}$ such that:
\begin{align*} 
  f(\vec{x}) = \frac{1}{2} \vec{x}^\dagger A \vec{x} - \vec{x}^\dagger \vec{b}
               + \text{const.} \quad.
\end{align*}
A vector $\vec{x}$ is conjugate to another vector $\vec{y}$ with respect to 
$f$ if and only if $\vec{x}^\dagger A \vec{y} = 0$. The gradient of the function is
\begin{align*}
  \nabla f(\vec{x}) = A \vec{x} - \vec{b}
\end{align*}
such that a stationary point $\vec{x}_*$ satisfies $A \vec{x}_* = \vec{b}$.
Given a basis consisting of $n$ mutually conjugate vectors 
$\vec{p}_n^\dagger A \vec{p}_m = \delta_{nm}$, we can expand $\vec{x}_*$ in
that basis $\vec{x}_* = \sum_n c_n \vec{p}_n$ with coefficients 
dependent only on the corresponding basis vectors:
\begin{align*}
  c_n = \frac{\vec{p}_n^\dagger \vec{b}}{\vec{p}_n^\dagger A \vec{p}_n}.
\end{align*}
This means we can pick a starting vector $\vec{p}_0$ and proceed to the minimum in
exactly $n$ steps by finding successive $\vec{p}_n$ that are conjugate to all
previous $\vec{p}_{0\dots n-1}$, where the step-size $c_n$ is uniquely determined 
by the current direction $\vec{p}_n$. This can be further improved on by choosing
specific $\vec{p}_n$
\begin{align*}
 \vec{p}_{n+1} = \vec{r}_{n+1} + \beta_n \vec{p}_n,
\end{align*}
where $\vec{r}_n = -\nabla f(\vec{x}_n)$ is the negative gradient and $\beta_n$
is a number defined below.
We make steps
\begin{align*}
 \vec{x}_{n+1} = \vec{x}_{n} + \alpha_n \vec{p}_n
\end{align*}
with 
\begin{align*}
 \vec{r}_{n+1} = -\nabla f(\vec{x}_{n+1}) = \vec{r}_n + \alpha_n A \vec{p}_n,
\end{align*}
where $\vec{p}_0 = \vec{r}_0 = \vec{b} - A \vec{x}_0$. Requiring
$\vec{r}_n^\dagger \vec{r}_m = \delta_{nm}$ and 
$\vec{p}_n^\dagger A \vec{p}_m = \delta_{nm}$ then result in
\begin{align*}
 \alpha_n = \frac{\vec{r}^\dagger_n \vec{r}_n}{\vec{p}_n^\dagger A \vec{p}_n}
 \quad \text{and} \quad
 \beta_n = \frac{\vec{r}^\dagger_{n+1} \vec{r}_{n+1}}{\vec{r}^\dagger_n \vec{r}_n}
\end{align*}
so that we only need to know $\vec{p}_n$ and $\vec{r}_n$ to calculate the next step.
The red line in the above illustration demonstrates this procedure.
We can modify this version of the conjugate-gradient method again so that it
can iteratively find the solution of \emph{approximately} quadratic problems. In this 
case, we must treat the function $f$ and its gradient $\nabla f$ as black boxes.
We can do this by obtaining an approximate $\alpha$, which is the only quantity 
requiring direct knowledge of $A$, by doing a line-search to find the minimum of 
$f(\vec{x}_n + \alpha \vec{p}_n)$. The algorithm, known as the non-linear 
conjugate gradient method, is then:
\begin{enumerate}
  \item Calculate $\vec{r}_n = -\nabla f(\vec{x}_n)$.
  \item Compute $\beta_{n-1}$.
  \item Calculate the next conjugate vector $\vec{p}_n = \vec{r}_n + \beta_{n-1} \vec{p}_{n-1}$.
  \item Use a line-search to find $\alpha_n = \arg \min_\alpha f(\vec{x}_n + \alpha \vec{p}_n)$.
  \item Set new position $\vec{x}_{n+1} = \vec{x}_n + \alpha_n \vec{p}_n$.
\end{enumerate}
The initial values are $\vec{p}_0 = \vec{r}_0 = -\nabla f(\vec{x}_0)$.
If $f$ is exactly quadratic and we ignore all numerical error, this algorithm
will find the minimum in $d$ iterations or less. With an approximately
quadratic $f$ and/or accounting for limited numerical precision, the vectors
$p_n$ will not be exactly conjugate to each other and errors will
accumulate. The algorithm must therefore be re-started at least every $d$
iterations. Note that there are other choices for $\beta_n$ that are
equivalent in the quadratic case, but result in different non-linear conjugate
gradient algorithms. The above choice is the one originally used by 
Fletcher and Reeves \cite{fletcher_function_1964}.

\section{Matrix elements for phi-4-theory}
\label{apn_phi4_matrix_elements}
The matrix elements of the operators needed to implement the $\phi^4$-theory
Hamiltonian using the site number-basis defined in section \ref{chap:phi4}
are as follows:

\begin{align*}
  \braket{s|\phi|t} = &\frac{1}{\sqrt{2}}\left[\delta_{s-1,t} \sqrt{(t+1)} \right.\\
                      &\left. + \delta_{s,t-1} \sqrt{(s+1)} \right],
\end{align*}
\begin{align*}      
  \braket{s|\phi^2|t} = &\frac{1}{2}\left[\delta_{s-1,t+1} \sqrt{(t+1)(t+2)} \right.\\
                        &+ \delta_{s,t} (2s + 1)\\
                        &\left. + \delta_{s+1,t-1} \sqrt{(s+1)(s+2)} \right],
\end{align*}
\begin{align*}
  \braket{s|\phi^3|t} = &\frac{1}{2\sqrt{2}}\left[
                        \delta_{s,t+3}\sqrt{(t+1)(t+2)(t+3)} \right.\\
                        &+\delta_{s,t+1} t \sqrt{(t+1)} \\
                        &+\delta_{s,t+1} ((t + 1) + (t + 2)) \sqrt{(t+1)} \\
                        &+\delta_{s+1,t} ((s + 1) + (s + 2)) \sqrt{(s+1)} \\
                        &+\delta_{s+1,t} s \sqrt{(s+1)} \\
                        &\left. +\delta_{s+3,t}\sqrt{(s+1)(s+2)(s+3)}  \right],
\end{align*}
\begin{align*}      
  \braket{s|\phi^4|t} = &\frac{1}{4}\left[
                        \delta_{s-1,t+3}\sqrt{(t+1)(t+2)(t+3)(t+4)} \right.\\
                        &+\delta_{s,t+2} (4t + 6) \sqrt{(t+1)(t+2)} \\
                        &+\delta_{s,t} (6t^2 + 6t + 3) \\
                        &+\delta_{s+2,t} (4s + 6) \sqrt{(s+1)(s+2)} \\
                        &\left. +\delta_{s+3,t-1}\sqrt{(s+1)(s+2)(s+3)(s+4)}  \right],
\end{align*}
\begin{align*}      
  \braket{s|\pi|t} = &\frac{\ic}{\sqrt{2}}\left[ \delta_{s,t+1} \sqrt{(t+1)} \right.\\
                     &\left. - \delta_{s+1,t} \sqrt{(s+1)} \right],
\end{align*}
\begin{align*}      
  \braket{s|\pi^2|t} = &\frac{1}{2} \left[ -\delta_{s-1,t+1} \sqrt{(t+1)(t+2)} \right.\\
                       &+ \delta_{s,t} (2s+1) \\
                       &\left. - \delta_{s+1,t-1} \sqrt{(s+1)(s+2)} \right].
\end{align*}


\begin{thebibliography}{51}%
\makeatletter
\providecommand \@ifxundefined [1]{%
 \@ifx{#1\undefined}
}%
\providecommand \@ifnum [1]{%
 \ifnum #1\expandafter \@firstoftwo
 \else \expandafter \@secondoftwo
 \fi
}%
\providecommand \@ifx [1]{%
 \ifx #1\expandafter \@firstoftwo
 \else \expandafter \@secondoftwo
 \fi
}%
\providecommand \natexlab [1]{#1}%
\providecommand \enquote  [1]{``#1''}%
\providecommand \bibnamefont  [1]{#1}%
\providecommand \bibfnamefont [1]{#1}%
\providecommand \citenamefont [1]{#1}%
\providecommand \href@noop [0]{\@secondoftwo}%
\providecommand \href [0]{\begingroup \@sanitize@url \@href}%
\providecommand \@href[1]{\@@startlink{#1}\@@href}%
\providecommand \@@href[1]{\endgroup#1\@@endlink}%
\providecommand \@sanitize@url [0]{\catcode `\\12\catcode `\$12\catcode
  `\&12\catcode `\#12\catcode `\^12\catcode `\_12\catcode `\%12\relax}%
\providecommand \@@startlink[1]{}%
\providecommand \@@endlink[0]{}%
\providecommand \url  [0]{\begingroup\@sanitize@url \@url }%
\providecommand \@url [1]{\endgroup\@href {#1}{\urlprefix }}%
\providecommand \urlprefix  [0]{URL }%
\providecommand \Eprint [0]{\href }%
\providecommand \doibase [0]{http://dx.doi.org/}%
\providecommand \selectlanguage [0]{\@gobble}%
\providecommand \bibinfo  [0]{\@secondoftwo}%
\providecommand \bibfield  [0]{\@secondoftwo}%
\providecommand \translation [1]{[#1]}%
\providecommand \BibitemOpen [0]{}%
\providecommand \bibitemStop [0]{}%
\providecommand \bibitemNoStop [0]{.\EOS\space}%
\providecommand \EOS [0]{\spacefactor3000\relax}%
\providecommand \BibitemShut  [1]{\csname bibitem#1\endcsname}%
\let\auto@bib@innerbib\@empty
\bibitem [{\citenamefont {Peskin}\ and\ \citenamefont
  {Schroeder}(1995)}]{peskin_introduction_1995}%
  \BibitemOpen
  \bibfield  {author} {\bibinfo {author} {\bibfnamefont {M.~E.}\ \bibnamefont
  {Peskin}}\ and\ \bibinfo {author} {\bibfnamefont {D.~V.}\ \bibnamefont
  {Schroeder}},\ }\href@noop {} {{\selectlanguage {english}\emph {\bibinfo
  {title} {An Introduction To Quantum Field Theory}}}}\ (\bibinfo  {publisher}
  {Westview Press},\ \bibinfo {year} {1995})\BibitemShut {NoStop}%
\bibitem [{\citenamefont {ATLAS-Collaboration}(2012)}]{atlas_observation_2012}%
  \BibitemOpen
  \bibfield  {author} {\bibinfo {author} {\bibnamefont {ATLAS-Collaboration}},\
  }\href {\doibase 10.1016/j.physletb.2012.08.020} {\bibfield  {journal}
  {\bibinfo  {journal} {Phys Lett B}\ }\textbf {\bibinfo {volume} {716}},\
  \bibinfo {pages} {1} (\bibinfo {year} {2012})}\BibitemShut {NoStop}%
\bibitem [{\citenamefont {Creutz}(1985)}]{creutz_quarks_1985}%
  \BibitemOpen
  \bibfield  {author} {\bibinfo {author} {\bibfnamefont {M.}~\bibnamefont
  {Creutz}},\ }\href@noop {} {{\selectlanguage {english}\emph {\bibinfo {title}
  {Quarks, Gluons and Lattices}}}}\ (\bibinfo  {publisher} {Cambridge
  University Press},\ \bibinfo {year} {1985})\BibitemShut {NoStop}%
\bibitem [{\citenamefont {von~der Linden}(1992)}]{von_der_linden_quantum_1992}%
  \BibitemOpen
  \bibfield  {author} {\bibinfo {author} {\bibfnamefont {W.}~\bibnamefont
  {von~der Linden}},\ }\href {\doibase 10.1016/0370-1573(92)90029-Y} {\bibfield
   {journal} {\bibinfo  {journal} {Phys. Rep.}\ }\textbf {\bibinfo {volume}
  {220}},\ \bibinfo {pages} {53} (\bibinfo {year} {1992})}\BibitemShut
  {NoStop}%
\bibitem [{\citenamefont {Dürr}\ \emph {et~al.}(2008)\citenamefont {Dürr},
  \citenamefont {Fodor}, \citenamefont {Frison}, \citenamefont {Hoelbling},
  \citenamefont {Hoffmann}, \citenamefont {Katz}, \citenamefont {Krieg},
  \citenamefont {Kurth}, \citenamefont {Lellouch}, \citenamefont {Lippert},
  \citenamefont {Szabo},\ and\ \citenamefont {Vulvert}}]{durr_ab_2008}%
  \BibitemOpen
  \bibfield  {author} {\bibinfo {author} {\bibfnamefont {S.}~\bibnamefont
  {Dürr}}, \bibinfo {author} {\bibfnamefont {Z.}~\bibnamefont {Fodor}},
  \bibinfo {author} {\bibfnamefont {J.}~\bibnamefont {Frison}}, \bibinfo
  {author} {\bibfnamefont {C.}~\bibnamefont {Hoelbling}}, \bibinfo {author}
  {\bibfnamefont {R.}~\bibnamefont {Hoffmann}}, \bibinfo {author}
  {\bibfnamefont {S.~D.}\ \bibnamefont {Katz}}, \bibinfo {author}
  {\bibfnamefont {S.}~\bibnamefont {Krieg}}, \bibinfo {author} {\bibfnamefont
  {T.}~\bibnamefont {Kurth}}, \bibinfo {author} {\bibfnamefont
  {L.}~\bibnamefont {Lellouch}}, \bibinfo {author} {\bibfnamefont
  {T.}~\bibnamefont {Lippert}}, \bibinfo {author} {\bibfnamefont {K.~K.}\
  \bibnamefont {Szabo}}, \ and\ \bibinfo {author} {\bibfnamefont
  {G.}~\bibnamefont {Vulvert}},\ }\href {\doibase 10.1126/science.1163233}
  {\bibfield  {journal} {\bibinfo  {journal} {Int S Techn Pol Inn}\ }\textbf
  {\bibinfo {volume} {322}},\ \bibinfo {pages} {1224} (\bibinfo {year}
  {2008})}\BibitemShut {NoStop}%
\bibitem [{\citenamefont {White}(1992)}]{white_density_1992}%
  \BibitemOpen
  \bibfield  {author} {\bibinfo {author} {\bibfnamefont {S.~R.}\ \bibnamefont
  {White}},\ }\href {\doibase 10.1103/PhysRevLett.69.2863} {\bibfield
  {journal} {\bibinfo  {journal} {Phys Rev Lett}\ }\textbf {\bibinfo {volume}
  {69}},\ \bibinfo {pages} {2863} (\bibinfo {year} {1992})}\BibitemShut
  {NoStop}%
\bibitem [{\citenamefont {Schollwöck}(2005)}]{schollwock_density-matrix_2005}%
  \BibitemOpen
  \bibfield  {author} {\bibinfo {author} {\bibfnamefont {U.}~\bibnamefont
  {Schollwöck}},\ }\href {\doibase 10.1103/RevModPhys.77.259} {\bibfield
  {journal} {\bibinfo  {journal} {Rev Mod Phys}\ }\textbf {\bibinfo {volume}
  {77}},\ \bibinfo {pages} {259} (\bibinfo {year} {2005})}\BibitemShut
  {NoStop}%
\bibitem [{\citenamefont {Eisert}\ \emph {et~al.}(2010)\citenamefont {Eisert},
  \citenamefont {Cramer},\ and\ \citenamefont
  {Plenio}}]{eisert_colloquium:_2010}%
  \BibitemOpen
  \bibfield  {author} {\bibinfo {author} {\bibfnamefont {J.}~\bibnamefont
  {Eisert}}, \bibinfo {author} {\bibfnamefont {M.}~\bibnamefont {Cramer}}, \
  and\ \bibinfo {author} {\bibfnamefont {M.~B.}\ \bibnamefont {Plenio}},\
  }\href {\doibase 10.1103/RevModPhys.82.277} {\bibfield  {journal} {\bibinfo
  {journal} {Rev Mod Phys}\ }\textbf {\bibinfo {volume} {82}},\ \bibinfo
  {pages} {277} (\bibinfo {year} {2010})}\BibitemShut {NoStop}%
\bibitem [{\citenamefont {Hastings}(2007)}]{hastings_area_2007}%
  \BibitemOpen
  \bibfield  {author} {\bibinfo {author} {\bibfnamefont {M.~B.}\ \bibnamefont
  {Hastings}},\ }\href {\doibase 10.1088/1742-5468/2007/08/P08024} {\bibfield
  {journal} {\bibinfo  {journal} {J Stat Mech-Theory E}\ }\textbf {\bibinfo
  {volume} {2007}},\ \bibinfo {pages} {P08024} (\bibinfo {year}
  {2007})}\BibitemShut {NoStop}%
\bibitem [{\citenamefont
  {Osborne}(2006{\natexlab{a}})}]{osborne_efficient_2006}%
  \BibitemOpen
  \bibfield  {author} {\bibinfo {author} {\bibfnamefont {T.~J.}\ \bibnamefont
  {Osborne}},\ }\href {\doibase 10.1103/PhysRevLett.97.157202} {\bibfield
  {journal} {\bibinfo  {journal} {Phys Rev Lett}\ }\textbf {\bibinfo {volume}
  {97}},\ \bibinfo {pages} {157202} (\bibinfo {year}
  {2006}{\natexlab{a}})}\BibitemShut {NoStop}%
\bibitem [{\citenamefont {de~Beaudrap}\ \emph {et~al.}(2010)\citenamefont
  {de~Beaudrap}, \citenamefont {Ohliger}, \citenamefont {Osborne},\ and\
  \citenamefont {Eisert}}]{de_beaudrap_solving_2010}%
  \BibitemOpen
  \bibfield  {author} {\bibinfo {author} {\bibfnamefont {N.}~\bibnamefont
  {de~Beaudrap}}, \bibinfo {author} {\bibfnamefont {M.}~\bibnamefont
  {Ohliger}}, \bibinfo {author} {\bibfnamefont {T.~J.}\ \bibnamefont
  {Osborne}}, \ and\ \bibinfo {author} {\bibfnamefont {J.}~\bibnamefont
  {Eisert}},\ }\href {\doibase 10.1103/PhysRevLett.105.060504} {\bibfield
  {journal} {\bibinfo  {journal} {Phys Rev Lett}\ }\textbf {\bibinfo {volume}
  {105}},\ \bibinfo {pages} {060504} (\bibinfo {year} {2010})}\BibitemShut
  {NoStop}%
\bibitem [{\citenamefont {Hastings}(2006)}]{hastings_solving_2006}%
  \BibitemOpen
  \bibfield  {author} {\bibinfo {author} {\bibfnamefont {M.~B.}\ \bibnamefont
  {Hastings}},\ }\href {\doibase 10.1103/PhysRevB.73.085115} {\bibfield
  {journal} {\bibinfo  {journal} {Phys Rev B}\ }\textbf {\bibinfo {volume}
  {73}},\ \bibinfo {pages} {085115} (\bibinfo {year} {2006})}\BibitemShut
  {NoStop}%
\bibitem [{\citenamefont {Masanes}(2009)}]{masanes_area_2009}%
  \BibitemOpen
  \bibfield  {author} {\bibinfo {author} {\bibfnamefont {L.}~\bibnamefont
  {Masanes}},\ }\href {\doibase 10.1103/PhysRevA.80.052104} {\bibfield
  {journal} {\bibinfo  {journal} {Phys Rev A}\ }\textbf {\bibinfo {volume}
  {80}},\ \bibinfo {pages} {052104} (\bibinfo {year} {2009})}\BibitemShut
  {NoStop}%
\bibitem [{\citenamefont {Haegeman}\ \emph {et~al.}(2011)\citenamefont
  {Haegeman}, \citenamefont {Cirac}, \citenamefont {Osborne}, \citenamefont
  {Pižorn}, \citenamefont {Verschelde},\ and\ \citenamefont
  {Verstraete}}]{haegeman_time-dependent_2011}%
  \BibitemOpen
  \bibfield  {author} {\bibinfo {author} {\bibfnamefont {J.}~\bibnamefont
  {Haegeman}}, \bibinfo {author} {\bibfnamefont {J.~I.}\ \bibnamefont {Cirac}},
  \bibinfo {author} {\bibfnamefont {T.~J.}\ \bibnamefont {Osborne}}, \bibinfo
  {author} {\bibfnamefont {I.}~\bibnamefont {Pižorn}}, \bibinfo {author}
  {\bibfnamefont {H.}~\bibnamefont {Verschelde}}, \ and\ \bibinfo {author}
  {\bibfnamefont {F.}~\bibnamefont {Verstraete}},\ }\href {\doibase
  10.1103/PhysRevLett.107.070601} {\bibfield  {journal} {\bibinfo  {journal}
  {Phys Rev Lett}\ }\textbf {\bibinfo {volume} {107}},\ \bibinfo {pages}
  {070601} (\bibinfo {year} {2011})}\BibitemShut {NoStop}%
\bibitem [{\citenamefont {Haegeman}\ \emph
  {et~al.}(2012{\natexlab{a}})\citenamefont {Haegeman}, \citenamefont {Pirvu},
  \citenamefont {Weir}, \citenamefont {Cirac}, \citenamefont {Osborne},
  \citenamefont {Verschelde},\ and\ \citenamefont
  {Verstraete}}]{haegeman_variational_2012}%
  \BibitemOpen
  \bibfield  {author} {\bibinfo {author} {\bibfnamefont {J.}~\bibnamefont
  {Haegeman}}, \bibinfo {author} {\bibfnamefont {B.}~\bibnamefont {Pirvu}},
  \bibinfo {author} {\bibfnamefont {D.~J.}\ \bibnamefont {Weir}}, \bibinfo
  {author} {\bibfnamefont {J.~I.}\ \bibnamefont {Cirac}}, \bibinfo {author}
  {\bibfnamefont {T.~J.}\ \bibnamefont {Osborne}}, \bibinfo {author}
  {\bibfnamefont {H.}~\bibnamefont {Verschelde}}, \ and\ \bibinfo {author}
  {\bibfnamefont {F.}~\bibnamefont {Verstraete}},\ }\href {\doibase
  10.1103/PhysRevB.85.100408} {\bibfield  {journal} {\bibinfo  {journal} {Phys
  Rev B}\ }\textbf {\bibinfo {volume} {85}},\ \bibinfo {pages} {100408}
  (\bibinfo {year} {2012}{\natexlab{a}})}\BibitemShut {NoStop}%
\bibitem [{\citenamefont {Chang}(1976)}]{chang_existence_1976}%
  \BibitemOpen
  \bibfield  {author} {\bibinfo {author} {\bibfnamefont {S.-J.}\ \bibnamefont
  {Chang}},\ }\href {\doibase 10.1103/PhysRevD.13.2778} {\bibfield  {journal}
  {\bibinfo  {journal} {Phys Rev D}\ }\textbf {\bibinfo {volume} {13}},\
  \bibinfo {pages} {2778} (\bibinfo {year} {1976})}\BibitemShut {NoStop}%
\bibitem [{\citenamefont {Simon}\ and\ \citenamefont
  {Griffiths}(1973)}]{simon_phi4_2_1973}%
  \BibitemOpen
  \bibfield  {author} {\bibinfo {author} {\bibfnamefont {B.}~\bibnamefont
  {Simon}}\ and\ \bibinfo {author} {\bibfnamefont {R.~B.}\ \bibnamefont
  {Griffiths}},\ }\href@noop {} {\bibfield  {journal} {\bibinfo  {journal}
  {Commun Math Phys}\ }\textbf {\bibinfo {volume} {33}},\ \bibinfo {pages}
  {145–164} (\bibinfo {year} {1973})}\BibitemShut {NoStop}%
\bibitem [{\citenamefont {Sugihara}(2004)}]{sugihara_density_2004}%
  \BibitemOpen
  \bibfield  {author} {\bibinfo {author} {\bibfnamefont {T.}~\bibnamefont
  {Sugihara}},\ }\href {\doibase 10.1088/1126-6708/2004/05/007} {\bibfield
  {journal} {\bibinfo  {journal} {J High Energy Phys}\ }\textbf {\bibinfo
  {volume} {2004}},\ \bibinfo {pages} {007} (\bibinfo {year}
  {2004})}\BibitemShut {NoStop}%
\bibitem [{\citenamefont {Weir}(2010)}]{weir_studying_2010}%
  \BibitemOpen
  \bibfield  {author} {\bibinfo {author} {\bibfnamefont {D.~J.}\ \bibnamefont
  {Weir}},\ }\href {\doibase 10.1103/PhysRevD.82.025003} {\bibfield  {journal}
  {\bibinfo  {journal} {Phys Rev D}\ }\textbf {\bibinfo {volume} {82}},\
  \bibinfo {pages} {025003} (\bibinfo {year} {2010})}\BibitemShut {NoStop}%
\bibitem [{\citenamefont {Loinaz}\ and\ \citenamefont
  {Willey}(1998)}]{loinaz_monte_1998}%
  \BibitemOpen
  \bibfield  {author} {\bibinfo {author} {\bibfnamefont {W.}~\bibnamefont
  {Loinaz}}\ and\ \bibinfo {author} {\bibfnamefont {R.~S.}\ \bibnamefont
  {Willey}},\ }\href {\doibase 10.1103/PhysRevD.58.076003} {\bibfield
  {journal} {\bibinfo  {journal} {Phys Rev D}\ }\textbf {\bibinfo {volume}
  {58}},\ \bibinfo {pages} {076003} (\bibinfo {year} {1998})}\BibitemShut
  {NoStop}%
\bibitem [{\citenamefont {Schaich}\ and\ \citenamefont
  {Loinaz}(2009)}]{schaich_improved_2009}%
  \BibitemOpen
  \bibfield  {author} {\bibinfo {author} {\bibfnamefont {D.}~\bibnamefont
  {Schaich}}\ and\ \bibinfo {author} {\bibfnamefont {W.}~\bibnamefont
  {Loinaz}},\ }\href {\doibase 10.1103/PhysRevD.79.056008} {\bibfield
  {journal} {\bibinfo  {journal} {Phys Rev D}\ }\textbf {\bibinfo {volume}
  {79}},\ \bibinfo {pages} {056008} (\bibinfo {year} {2009})}\BibitemShut
  {NoStop}%
\bibitem [{\citenamefont {De}\ \emph {et~al.}(2005)\citenamefont {De},
  \citenamefont {Harindranath}, \citenamefont {Maiti},\ and\ \citenamefont
  {Sinha}}]{de_investigations_2005}%
  \BibitemOpen
  \bibfield  {author} {\bibinfo {author} {\bibfnamefont {A.~K.}\ \bibnamefont
  {De}}, \bibinfo {author} {\bibfnamefont {A.}~\bibnamefont {Harindranath}},
  \bibinfo {author} {\bibfnamefont {J.}~\bibnamefont {Maiti}}, \ and\ \bibinfo
  {author} {\bibfnamefont {T.}~\bibnamefont {Sinha}},\ }\href {\doibase
  10.1103/PhysRevD.72.094503} {\bibfield  {journal} {\bibinfo  {journal} {Phys
  Rev D}\ }\textbf {\bibinfo {volume} {72}},\ \bibinfo {pages} {094503}
  (\bibinfo {year} {2005})}\BibitemShut {NoStop}%
\bibitem [{\citenamefont {Milsted}()}]{milsted_evomps_????}%
  \BibitemOpen
  \bibfield  {author} {\bibinfo {author} {\bibfnamefont {A.}~\bibnamefont
  {Milsted}},\ }\href {https://github.com/amilsted/evoMPS} {\enquote {\bibinfo
  {title} {{evoMPS}},}\ }\bibinfo {howpublished} {GitHub (BSD
  License)}\BibitemShut {NoStop}%
\bibitem [{\citenamefont {Brydges}\ \emph {et~al.}(1983)\citenamefont
  {Brydges}, \citenamefont {Fröhlich},\ and\ \citenamefont
  {Sokal}}]{brydges_new_1983}%
  \BibitemOpen
  \bibfield  {author} {\bibinfo {author} {\bibfnamefont {D.~C.}\ \bibnamefont
  {Brydges}}, \bibinfo {author} {\bibfnamefont {J.}~\bibnamefont {Fröhlich}},
  \ and\ \bibinfo {author} {\bibfnamefont {A.~D.}\ \bibnamefont {Sokal}},\
  }\href@noop {} {\bibfield  {journal} {\bibinfo  {journal} {Commun Math Phys}\
  }\textbf {\bibinfo {volume} {91}},\ \bibinfo {pages} {141} (\bibinfo {year}
  {1983})}\BibitemShut {NoStop}%
\bibitem [{\citenamefont {Jordan}\ \emph {et~al.}(2011)\citenamefont {Jordan},
  \citenamefont {Lee},\ and\ \citenamefont {Preskill}}]{Jordan_quantum_2011}%
  \BibitemOpen
  \bibfield  {author} {\bibinfo {author} {\bibfnamefont {S.~P.}\ \bibnamefont
  {Jordan}}, \bibinfo {author} {\bibfnamefont {K.~S.~M.}\ \bibnamefont {Lee}},
  \ and\ \bibinfo {author} {\bibfnamefont {J.}~\bibnamefont {Preskill}},\
  }\href@noop {} {\bibfield  {journal} {\bibinfo  {journal} {{arXiv}}\ }
  (\bibinfo {year} {2011})},\ \Eprint {http://arxiv.org/abs/1112.4833}
  {1112.4833} \BibitemShut {NoStop}%
\bibitem [{\citenamefont {Jackiw}\ and\ \citenamefont
  {Templeton}(1981)}]{jackiw_how_1981}%
  \BibitemOpen
  \bibfield  {author} {\bibinfo {author} {\bibfnamefont {R.}~\bibnamefont
  {Jackiw}}\ and\ \bibinfo {author} {\bibfnamefont {S.}~\bibnamefont
  {Templeton}},\ }\href {\doibase 10.1103/PhysRevD.23.2291} {\bibfield
  {journal} {\bibinfo  {journal} {Phys Rev D}\ }\textbf {\bibinfo {volume}
  {23}},\ \bibinfo {pages} {2291} (\bibinfo {year} {1981})}\BibitemShut
  {NoStop}%
\bibitem [{\citenamefont {Sachdev}(2011)}]{sachdev_quantum_2011}%
  \BibitemOpen
  \bibfield  {author} {\bibinfo {author} {\bibfnamefont {S.}~\bibnamefont
  {Sachdev}},\ }\href@noop {} {\emph {\bibinfo {title} {Quantum Phase
  Transitions}}},\ \bibinfo {edition} {2nd}\ ed.\ (\bibinfo  {publisher}
  {Cambridge University Press},\ \bibinfo {year} {2011})\BibitemShut {NoStop}%
\bibitem [{\citenamefont {Kleinert}\ and\ \citenamefont
  {Schulte-Frohlinde}(2001)}]{kleinert_critical_2001}%
  \BibitemOpen
  \bibfield  {author} {\bibinfo {author} {\bibfnamefont {H.}~\bibnamefont
  {Kleinert}}\ and\ \bibinfo {author} {\bibfnamefont {V.}~\bibnamefont
  {Schulte-Frohlinde}},\ }\href@noop {} {\emph {\bibinfo {title} {Critical
  Properties of Phi-4-theories}}}\ (\bibinfo  {publisher} {World Scientific
  Publishing Co Pte Ltd},\ \bibinfo {year} {2001})\BibitemShut {NoStop}%
\bibitem [{\citenamefont {Haegeman}\ \emph
  {et~al.}(2012{\natexlab{b}})\citenamefont {Haegeman}, \citenamefont
  {Mariën}, \citenamefont {Osborne},\ and\ \citenamefont
  {Verstraete}}]{haegeman_geometry_2012}%
  \BibitemOpen
  \bibfield  {author} {\bibinfo {author} {\bibfnamefont {J.}~\bibnamefont
  {Haegeman}}, \bibinfo {author} {\bibfnamefont {M.}~\bibnamefont {Mariën}},
  \bibinfo {author} {\bibfnamefont {T.~J.}\ \bibnamefont {Osborne}}, \ and\
  \bibinfo {author} {\bibfnamefont {F.}~\bibnamefont {Verstraete}},\
  }\href@noop {} {\bibfield  {journal} {\bibinfo  {journal} {{arXiv}}\ }
  (\bibinfo {year} {2012}{\natexlab{b}})},\ \Eprint
  {http://arxiv.org/abs/1210.7710} {1210.7710} \BibitemShut {NoStop}%
\bibitem [{\citenamefont {Absil}\ \emph {et~al.}(2009)\citenamefont {Absil},
  \citenamefont {Mahony},\ and\ \citenamefont
  {Sepulchre}}]{absil_optimization_2009}%
  \BibitemOpen
  \bibfield  {author} {\bibinfo {author} {\bibfnamefont {P.-A.}\ \bibnamefont
  {Absil}}, \bibinfo {author} {\bibfnamefont {R.}~\bibnamefont {Mahony}}, \
  and\ \bibinfo {author} {\bibfnamefont {R.}~\bibnamefont {Sepulchre}},\
  }\href@noop {} {{\selectlanguage {english}\emph {\bibinfo {title}
  {Optimization Algorithms on Matrix Manifolds}}}}\ (\bibinfo  {publisher}
  {Princeton University Press},\ \bibinfo {year} {2009})\BibitemShut {NoStop}%
\bibitem [{\citenamefont {Pollmann}\ \emph {et~al.}(2009)\citenamefont
  {Pollmann}, \citenamefont {Mukerjee}, \citenamefont {Turner},\ and\
  \citenamefont {Moore}}]{pollmann_theory_2009}%
  \BibitemOpen
  \bibfield  {author} {\bibinfo {author} {\bibfnamefont {F.}~\bibnamefont
  {Pollmann}}, \bibinfo {author} {\bibfnamefont {S.}~\bibnamefont {Mukerjee}},
  \bibinfo {author} {\bibfnamefont {A.~M.}\ \bibnamefont {Turner}}, \ and\
  \bibinfo {author} {\bibfnamefont {J.~E.}\ \bibnamefont {Moore}},\ }\href
  {\doibase 10.1103/PhysRevLett.102.255701} {\bibfield  {journal} {\bibinfo
  {journal} {Phys Rev Lett}\ }\textbf {\bibinfo {volume} {102}},\ \bibinfo
  {pages} {255701} (\bibinfo {year} {2009})}\BibitemShut {NoStop}%
\bibitem [{\citenamefont {Latorre}\ \emph {et~al.}(2003)\citenamefont
  {Latorre}, \citenamefont {Rico},\ and\ \citenamefont
  {Vidal}}]{latorre_ground_2003}%
  \BibitemOpen
  \bibfield  {author} {\bibinfo {author} {\bibfnamefont {J.~I.}\ \bibnamefont
  {Latorre}}, \bibinfo {author} {\bibfnamefont {E.}~\bibnamefont {Rico}}, \
  and\ \bibinfo {author} {\bibfnamefont {G.}~\bibnamefont {Vidal}},\
  }\href@noop {} {\bibfield  {journal} {\bibinfo  {journal} {{arXiv}}\ }
  (\bibinfo {year} {2003})},\ \bibinfo {note} {{Quant Inf Comput} 4 (2004)
  48-92},\ \Eprint {http://arxiv.org/abs/quant-ph/0304098} {quant-ph/0304098}
  \BibitemShut {NoStop}%
\bibitem [{\citenamefont {Calabrese}\ and\ \citenamefont
  {Cardy}(2004)}]{calabrese_entanglement_2004}%
  \BibitemOpen
  \bibfield  {author} {\bibinfo {author} {\bibfnamefont {P.}~\bibnamefont
  {Calabrese}}\ and\ \bibinfo {author} {\bibfnamefont {J.}~\bibnamefont
  {Cardy}},\ }\href {\doibase 10.1088/1742-5468/2004/06/P06002} {\bibfield
  {journal} {\bibinfo  {journal} {J Stat Mech-Theory E}\ }\textbf {\bibinfo
  {volume} {2004}},\ \bibinfo {pages} {P06002} (\bibinfo {year}
  {2004})}\BibitemShut {NoStop}%
\bibitem [{\citenamefont {Vidal}\ \emph {et~al.}(2003)\citenamefont {Vidal},
  \citenamefont {Latorre}, \citenamefont {Rico},\ and\ \citenamefont
  {Kitaev}}]{vidal_entanglement_2003}%
  \BibitemOpen
  \bibfield  {author} {\bibinfo {author} {\bibfnamefont {G.}~\bibnamefont
  {Vidal}}, \bibinfo {author} {\bibfnamefont {J.~I.}\ \bibnamefont {Latorre}},
  \bibinfo {author} {\bibfnamefont {E.}~\bibnamefont {Rico}}, \ and\ \bibinfo
  {author} {\bibfnamefont {A.}~\bibnamefont {Kitaev}},\ }\href {\doibase
  10.1103/PhysRevLett.90.227902} {\bibfield  {journal} {\bibinfo  {journal}
  {Phys Rev Lett}\ }\textbf {\bibinfo {volume} {90}},\ \bibinfo {pages}
  {227902} (\bibinfo {year} {2003})}\BibitemShut {NoStop}%
\bibitem [{\citenamefont {Tagliacozzo}\ \emph {et~al.}(2008)\citenamefont
  {Tagliacozzo}, \citenamefont {de~Oliveira}, \citenamefont {Iblisdir},\ and\
  \citenamefont {Latorre}}]{tagliacozzo_scaling_2008}%
  \BibitemOpen
  \bibfield  {author} {\bibinfo {author} {\bibfnamefont {L.}~\bibnamefont
  {Tagliacozzo}}, \bibinfo {author} {\bibfnamefont {T.~R.}\ \bibnamefont
  {de~Oliveira}}, \bibinfo {author} {\bibfnamefont {S.}~\bibnamefont
  {Iblisdir}}, \ and\ \bibinfo {author} {\bibfnamefont {J.~I.}\ \bibnamefont
  {Latorre}},\ }\href {\doibase 10.1103/PhysRevB.78.024410} {\bibfield
  {journal} {\bibinfo  {journal} {Phys Rev B}\ }\textbf {\bibinfo {volume}
  {78}},\ \bibinfo {pages} {024410} (\bibinfo {year} {2008})}\BibitemShut
  {NoStop}%
\bibitem [{\citenamefont {Press}\ \emph {et~al.}(2007)\citenamefont {Press},
  \citenamefont {Teukolsky}, \citenamefont {Vetterling},\ and\ \citenamefont
  {Flannery}}]{press_numerical_2007}%
  \BibitemOpen
  \bibfield  {author} {\bibinfo {author} {\bibfnamefont {W.}~\bibnamefont
  {Press}}, \bibinfo {author} {\bibfnamefont {S.}~\bibnamefont {Teukolsky}},
  \bibinfo {author} {\bibfnamefont {W.}~\bibnamefont {Vetterling}}, \ and\
  \bibinfo {author} {\bibfnamefont {B.}~\bibnamefont {Flannery}},\ }\href@noop
  {} {\emph {\bibinfo {title} {Numerical Recipes: The Art of Scientific
  Computing}}},\ \bibinfo {edition} {3rd}\ ed.\ (\bibinfo  {publisher}
  {Cambridge University Press},\ \bibinfo {address} {New York},\ \bibinfo
  {year} {2007})\BibitemShut {NoStop}%
\bibitem [{\citenamefont
  {Osborne}(2006{\natexlab{b}})}]{osborne_renormalisation-group_2006}%
  \BibitemOpen
  \bibfield  {author} {\bibinfo {author} {\bibfnamefont {T.~J.}\ \bibnamefont
  {Osborne}},\ }\href@noop {} {\bibfield  {journal} {\bibinfo  {journal}
  {{arXiv}}\ } (\bibinfo {year} {2006}{\natexlab{b}})},\ \Eprint
  {http://arxiv.org/abs/cond-mat/0605194} {cond-mat/0605194} \BibitemShut
  {NoStop}%
\bibitem [{\citenamefont {Rommer}\ and\ \citenamefont
  {Östlund}(1997)}]{rommer_class_1997}%
  \BibitemOpen
  \bibfield  {author} {\bibinfo {author} {\bibfnamefont {S.}~\bibnamefont
  {Rommer}}\ and\ \bibinfo {author} {\bibfnamefont {S.}~\bibnamefont
  {Östlund}},\ }\href {\doibase 10.1103/PhysRevB.55.2164} {\bibfield
  {journal} {\bibinfo  {journal} {Phys Rev B}\ }\textbf {\bibinfo {volume}
  {55}},\ \bibinfo {pages} {2164} (\bibinfo {year} {1997})}\BibitemShut
  {NoStop}%
\bibitem [{\citenamefont {Häuser}\ \emph {et~al.}(1995)\citenamefont
  {Häuser}, \citenamefont {Cassing}, \citenamefont {Peter},\ and\
  \citenamefont {Thoma}}]{hauser_connected_1995}%
  \BibitemOpen
  \bibfield  {author} {\bibinfo {author} {\bibfnamefont {J.~M.}\ \bibnamefont
  {Häuser}}, \bibinfo {author} {\bibfnamefont {W.}~\bibnamefont {Cassing}},
  \bibinfo {author} {\bibfnamefont {A.}~\bibnamefont {Peter}}, \ and\ \bibinfo
  {author} {\bibfnamefont {M.~H.}\ \bibnamefont {Thoma}},\ }\href {\doibase
  10.1007/BF01292336} {\bibfield  {journal} {\bibinfo  {journal} {Z Phys
  A-Hadron Nuc}\ }\textbf {\bibinfo {volume} {353}},\ \bibinfo {pages} {301}
  (\bibinfo {year} {1995})}\BibitemShut {NoStop}%
\bibitem [{\citenamefont {Ji}\ \emph {et~al.}(2002)\citenamefont {Ji},
  \citenamefont {Kim}, \citenamefont {Min},\ and\ \citenamefont
  {Vinnikov}}]{ji_canonical_2002}%
  \BibitemOpen
  \bibfield  {author} {\bibinfo {author} {\bibfnamefont {C.-R.}\ \bibnamefont
  {Ji}}, \bibinfo {author} {\bibfnamefont {J.-I.}\ \bibnamefont {Kim}},
  \bibinfo {author} {\bibfnamefont {D.-P.}\ \bibnamefont {Min}}, \ and\
  \bibinfo {author} {\bibfnamefont {A.~V.}\ \bibnamefont {Vinnikov}},\
  }\href@noop {} {\bibfield  {journal} {\bibinfo  {journal} {{arXiv}}\ }
  (\bibinfo {year} {2002})},\ \Eprint {http://arxiv.org/abs/hep-ph/0204114}
  {hep-ph/0204114} \BibitemShut {NoStop}%
\bibitem [{\citenamefont {Lee}(1998)}]{lee_introduction_1998}%
  \BibitemOpen
  \bibfield  {author} {\bibinfo {author} {\bibfnamefont {D.}~\bibnamefont
  {Lee}},\ }\href {\doibase 10.1016/S0370-2693(98)01010-7} {\bibfield
  {journal} {\bibinfo  {journal} {Phys Lett B}\ }\textbf {\bibinfo {volume}
  {439}},\ \bibinfo {pages} {85} (\bibinfo {year} {1998})}\BibitemShut
  {NoStop}%
\bibitem [{\citenamefont {Marrero}\ \emph {et~al.}(1999)\citenamefont
  {Marrero}, \citenamefont {Roura},\ and\ \citenamefont
  {Lee}}]{marrero_non-perturbative_1999}%
  \BibitemOpen
  \bibfield  {author} {\bibinfo {author} {\bibfnamefont {P.~J.}\ \bibnamefont
  {Marrero}}, \bibinfo {author} {\bibfnamefont {E.~A.}\ \bibnamefont {Roura}},
  \ and\ \bibinfo {author} {\bibfnamefont {D.}~\bibnamefont {Lee}},\ }\href
  {\doibase 10.1016/S0370-2693(99)01341-6} {\bibfield  {journal} {\bibinfo
  {journal} {Phys Lett B}\ }\textbf {\bibinfo {volume} {471}},\ \bibinfo
  {pages} {45} (\bibinfo {year} {1999})}\BibitemShut {NoStop}%
\bibitem [{\citenamefont {Bender}\ \emph {et~al.}(1993)\citenamefont {Bender},
  \citenamefont {Pinsky},\ and\ \citenamefont {van~de
  Sande}}]{bender_spontaneous_1993}%
  \BibitemOpen
  \bibfield  {author} {\bibinfo {author} {\bibfnamefont {C.~M.}\ \bibnamefont
  {Bender}}, \bibinfo {author} {\bibfnamefont {S.}~\bibnamefont {Pinsky}}, \
  and\ \bibinfo {author} {\bibfnamefont {B.}~\bibnamefont {van~de Sande}},\
  }\href {\doibase 10.1103/PhysRevD.48.816} {\bibfield  {journal} {\bibinfo
  {journal} {Phys Rev D}\ }\textbf {\bibinfo {volume} {48}},\ \bibinfo {pages}
  {816} (\bibinfo {year} {1993})}\BibitemShut {NoStop}%
\bibitem [{\citenamefont {Funke}\ \emph {et~al.}(1987)\citenamefont {Funke},
  \citenamefont {Kaulfuss},\ and\ \citenamefont
  {Kümmel}}]{funke_approaching_1987}%
  \BibitemOpen
  \bibfield  {author} {\bibinfo {author} {\bibfnamefont {M.}~\bibnamefont
  {Funke}}, \bibinfo {author} {\bibfnamefont {U.}~\bibnamefont {Kaulfuss}}, \
  and\ \bibinfo {author} {\bibfnamefont {H.}~\bibnamefont {Kümmel}},\ }\href
  {\doibase 10.1103/PhysRevD.35.621} {\bibfield  {journal} {\bibinfo  {journal}
  {Phys Rev D}\ }\textbf {\bibinfo {volume} {35}},\ \bibinfo {pages} {621}
  (\bibinfo {year} {1987})}\BibitemShut {NoStop}%
\bibitem [{\citenamefont {Sugihara}(1998)}]{sugihara_variational_1998}%
  \BibitemOpen
  \bibfield  {author} {\bibinfo {author} {\bibfnamefont {T.}~\bibnamefont
  {Sugihara}},\ }\href {\doibase 10.1103/PhysRevD.57.7373} {\bibfield
  {journal} {\bibinfo  {journal} {Phys Rev D}\ }\textbf {\bibinfo {volume}
  {57}},\ \bibinfo {pages} {7373} (\bibinfo {year} {1998})}\BibitemShut
  {NoStop}%
\bibitem [{\citenamefont {Harindranath}\ and\ \citenamefont
  {Vary}(1987)}]{harindranath_solving_1987}%
  \BibitemOpen
  \bibfield  {author} {\bibinfo {author} {\bibfnamefont {A.}~\bibnamefont
  {Harindranath}}\ and\ \bibinfo {author} {\bibfnamefont {J.~P.}\ \bibnamefont
  {Vary}},\ }\href {\doibase 10.1103/PhysRevD.36.1141} {\bibfield  {journal}
  {\bibinfo  {journal} {Phys Rev D}\ }\textbf {\bibinfo {volume} {36}},\
  \bibinfo {pages} {1141} (\bibinfo {year} {1987})}\BibitemShut {NoStop}%
\bibitem [{\citenamefont {Harindranath}\ and\ \citenamefont
  {Vary}(1988)}]{harindranath_stability_1988}%
  \BibitemOpen
  \bibfield  {author} {\bibinfo {author} {\bibfnamefont {A.}~\bibnamefont
  {Harindranath}}\ and\ \bibinfo {author} {\bibfnamefont {J.~P.}\ \bibnamefont
  {Vary}},\ }\href {\doibase 10.1103/PhysRevD.37.1076} {\bibfield  {journal}
  {\bibinfo  {journal} {Phys Rev D}\ }\textbf {\bibinfo {volume} {37}},\
  \bibinfo {pages} {1076} (\bibinfo {year} {1988})}\BibitemShut {NoStop}%
\bibitem [{\citenamefont {Hansen}\ \emph {et~al.}(2002)\citenamefont {Hansen},
  \citenamefont {Chanfray}, \citenamefont {Davesne},\ and\ \citenamefont
  {Schuck}}]{hansen_random_2002}%
  \BibitemOpen
  \bibfield  {author} {\bibinfo {author} {\bibfnamefont {H.}~\bibnamefont
  {Hansen}}, \bibinfo {author} {\bibfnamefont {G.}~\bibnamefont {Chanfray}},
  \bibinfo {author} {\bibfnamefont {D.}~\bibnamefont {Davesne}}, \ and\
  \bibinfo {author} {\bibfnamefont {P.}~\bibnamefont {Schuck}},\ }\href
  {\doibase 10.1140/epja/i2002-10023-y} {\bibfield  {journal} {\bibinfo
  {journal} {Eur Phys J A}\ }\textbf {\bibinfo {volume} {14}},\ \bibinfo
  {pages} {397} (\bibinfo {year} {2002})}\BibitemShut {NoStop}%
\bibitem [{\citenamefont {Polley}\ and\ \citenamefont
  {Ritschel}(1989)}]{polley_second-order_1989}%
  \BibitemOpen
  \bibfield  {author} {\bibinfo {author} {\bibfnamefont {L.}~\bibnamefont
  {Polley}}\ and\ \bibinfo {author} {\bibfnamefont {U.}~\bibnamefont
  {Ritschel}},\ }\href {\doibase 10.1016/0370-2693(89)90189-5} {\bibfield
  {journal} {\bibinfo  {journal} {Phys Lett B}\ }\textbf {\bibinfo {volume}
  {221}},\ \bibinfo {pages} {44} (\bibinfo {year} {1989})}\BibitemShut
  {NoStop}%
\bibitem [{\citenamefont {Alexandrov}(2007)}]{cg_illustration}%
  \BibitemOpen
  \bibfield  {author} {\bibinfo {author} {\bibfnamefont {O.}~\bibnamefont
  {Alexandrov}},\ }\href
  {http://commons.wikimedia.org/wiki/File:Conjugate_gradient_illustration.svg}
  {\enquote {\bibinfo {title} {Conjugate gradient illustration (public
  domain)},}\ } (\bibinfo {year} {2007})\BibitemShut {NoStop}%
\bibitem [{\citenamefont {Fletcher}\ and\ \citenamefont
  {Reeves}(1964)}]{fletcher_function_1964}%
  \BibitemOpen
  \bibfield  {author} {\bibinfo {author} {\bibfnamefont {R.}~\bibnamefont
  {Fletcher}}\ and\ \bibinfo {author} {\bibfnamefont {C.~M.}\ \bibnamefont
  {Reeves}},\ }\href {\doibase 10.1093/comjnl/7.2.149} {\bibfield  {journal}
  {\bibinfo  {journal} {Comput. J}\ }\textbf {\bibinfo {volume} {7}},\ \bibinfo
  {pages} {149} (\bibinfo {year} {1964})}\BibitemShut {NoStop}%
\end{thebibliography}
\end{document}